\documentclass[12pt,a4paper]{article} % onecolumn (second format)

\RequirePackage{fix-cm}
\usepackage{authblk}
\usepackage{enumitem}
\usepackage{amsmath}
\usepackage{booktabs}
\usepackage{floatrow}
\DeclareFloatVCode{somespace}{\vspace{1.667\baselineskip}}
\usepackage{booktabs} 
\usepackage[utf8]{inputenc}
\usepackage{xcolor}
\usepackage{textcomp}
\usepackage{csquotes}
\usepackage[numbers,sort]{natbib}
\usepackage{graphicx}
\usepackage{xltabular}
 \usepackage{acronym}
\usepackage{graphicx}
\usepackage{hyperref}
\usepackage{rotating}
\usepackage{pgfplots}
\usepackage{multirow}
\usepackage{tikz}
\usepackage{lscape}

\usetikzlibrary{shapes,arrows}
\usetikzlibrary{intersections}
\usepackage[english]{babel}
\usepackage{comment}

\usepackage{float}
\floatstyle{plaintop}
\restylefloat{table}

 \newacro {SLR}[SLR]{systematic literature review}
 \newacro {SMR}[SMR]{systematic mapping review}
 
\newacro {ZSL}[ZSL]{zero-shot learning}
\newacro {NLP}[NLP]{natural language processing}
\newacro {ML}[ML]{machine learning}
\newacro {SO}[SO]{Stack Overflow}
\newacro {LDA} [LDA] {Latent Dirichlet Allocation} 
\newacro {OSS}[OSS]{open-source software}
\newacro {NLI}[NLI]{natural language inference}
 \newacro {GBT}[GBT]{gradient boosting tree}
 \newacro {SVM}[SVM]{support vector machine}
 \newacro{LLM}{large language model}
 \newacro{bLLM}{larger language model}
\newacro{FSL}{few-shot learning}
\newacro{sLLM}{smaller language model}
\newacro{TARS}{task-aware representation of sentences}

\begin {document}

\title{Sentiment analysis for software engineering: How far can \acf{ZSL} go?}

%\titlerunning{Short form of title} % if too long for running head

\author{Reem Alfayez \and
Manal Binkhonain %etc.
}

\affil{Department of Software Engineering, College of Computer and Information Sciences,\\
King Saud University, P.O. Box 51178, Riyadh 11543, Saudi Arabia\\
\texttt{reealfayez@ksu.edu.sa}}

\date{Received: date / Accepted: date}
% The correct dates will be entered by the editor

\maketitle
 \acresetall
\begin{abstract}
\textbf{Context:}
Sentiment analysis in software engineering focuses on understanding emotions expressed in software artifacts. Previous research highlighted the limitations of applying general off-the-shelf sentiment analysis tools within the software engineering domain and indicated the need for specialized tools tailored to various software engineering contexts. The development of such tools heavily relies on supervised machine learning techniques that necessitate annotated datasets. Acquiring such datasets is a substantial challenge, as it requires domain-specific expertise and significant effort. \textbf{Objective}: This study explores the potential of \acf{ZSL} to address the scarcity of annotated datasets in sentiment analysis within software engineering \textbf{Method:} We conducted an empirical experiment to evaluate the performance of various \ac{ZSL} techniques, including embedding-based, \ac{NLI}-based, \ac{TARS}-based, and generative-based \ac{ZSL} techniques. We assessed the performance of these techniques under different labels setups to examine the impact of label configurations. Additionally, we compared the results of the \ac{ZSL} techniques with state-of-the-art fine-tuned transformer-based models. Finally, we performed an error analysis to identify the primary causes of misclassifications. \textbf{Results:} Our findings demonstrate that \ac{ZSL} techniques, particularly those combining expert-curated labels with embedding-based or generative-based models, can achieve macro-F1 scores comparable to fine-tuned transformer-based models. The error analysis revealed that subjectivity in annotation and polar facts are the main contributors to \ac{ZSL} misclassifications.
\textbf{Conclusion:} This study demonstrates the potential of \ac{ZSL} for sentiment analysis in software engineering. \ac{ZSL} can provide a solution to the challenge of annotated dataset scarcity by reducing reliance on annotated dataset.
\end{abstract}

 \acresetall

\section{Introduction} 

Over the years, sentiment analysis has evolved as a powerful tool for extracting subjective information from text data \cite{zhang2020sentiment}. In software engineering, it provides insights into software development and usage by analyzing app reviews, developer communications, discussions on technical Q\&A websites, and more \cite{zhang2020sentiment, zhang2023revisiting, obaidi2021development, lin2018sentiment, sajadi2023towards, novielli2020can, uddin2019automatic, calefato2018sentiment}. 

Utilizing sentiment analysis within the software engineering domain presents significant challenges. General-purpose sentiment analysis tools often demonstrate suboptimal performance when used in software engineering contexts \cite{jongeling2015choosing,tourani2014monitoring}. This limitation has led researchers to develop specialized tools tailored to the unique characteristics of software engineering contexts. Despite these efforts, it has been observed that even sentiment analysis tools that perform well in one software engineering context lack generalizability across different contexts, where they underperform in the new contexts. The context-bound limitation highlights the need for context-specific tools \cite{novielli2018benchmark,lin2018sentiment,obaidi2021development}.

Developing context-based tools is not an easy feat, as most sentiment analysis tools in software engineering are based on supervised machine learning techniques that frame sentiment analysis tasks as text classification problems. These tools are heavily reliant on annotated datasets for training, which is costly, time-consuming, error-prone, and requires domain-specific expertise to obtain \cite{novielli2018benchmark,lin2018sentiment,zhang2023revisiting,lin2022opinion}.

This reliance on annotated datasets is particularly problematic due to the context-bound limitation that requires the development of 
 separate models and datasets for each specific context \cite{novielli2018benchmark,lin2018sentiment}.
 The need to curate annotated dataset for training in each context further intensifies the challenge of datasets scarcity and complicates the development of effective sentiment analysis models \cite{novielli2018benchmark,lin2018sentiment,zhang2023revisiting}.
 
\Ac{ZSL} is a promising approach that has the potential to address the challenge of requiring context-specific training data for classification tasks. A \ac{ZSL}-based model can classify data without prior exposure to specific labels, where it instead relies on its understanding of relationships between words, phrases, and concepts to make classifications. In the context of \ac{ZSL}, a model generates predictions for tasks it has not been explicitly trained on by leveraging data from other, related tasks to aid its learning process. The model utilizes knowledge acquired through pre-training on other datasets and transfers relevant information to the new classification task \cite{tunstall2022natural,alammar2024hands}.

This study explores the potential of \ac{ZSL} for sentiment analysis in software engineering by evaluating embedding-based, \ac{NLI}-based, \ac{TARS}-based, and generative-based \ac{ZSL} techniques across various contexts. We assess the impact of label configurations on model performance and compare the best-performing \ac{ZSL} models with state-of-the-art fine-tuned transformer models. An error analysis is also conducted to understand misclassification causes.

The rest of the paper is organized as follows: Section \ref{Background} summarizes \ac{ZSL} for text classification. Section \ref{Related work} reviews related studies. Section \ref{Study setup} describes the study setup. Section \ref{Results} presents the results, and Section \ref{Discussion} discusses them. Section \ref{Threats to validity} outlines potential validity threats, and Section \ref{Conclusion} concludes the paper.

\section{\Acf{ZSL} text classification} \label{Background}
\Ac{ZSL} text classification is a \ac{NLP} approach that enables models to classify text into unseen classes during training. As opposed to traditional supervised learning, which requires labeled data for each class, \ac{ZSL} leverages transfer learning and semantic understanding to predict previously unseen classes. This capability is particularly useful when labeled data is scarce \cite{alammar2024hands,tunstall2022natural}. \Ac{ZSL} text classification can be achieved through four main techniques: embedding-based, \ac{NLI}-based, \ac{TARS}-based, and generative-based techniques. Below is a brief description of each technique.

\subsection{Embedding-based \ac{ZSL}}
 Embedding-based \ac{ZSL} text classification uses word embeddings to measure the semantic similarity between input text and potential class labels \cite{veeranna2016using}. While the original approach \cite{veeranna2016using} relied on skip-gram static embeddings, we opted for transformer-based \acp{LLM} embeddings, as these models generate contextual word embeddings that capture both the syntactic and semantic properties of words, along with their context \cite{alammar2024hands,tunstall2022natural,alhoshan2023zero}.

\begin{figure} [htbp]
\center
 \includegraphics[width=.85\textwidth]{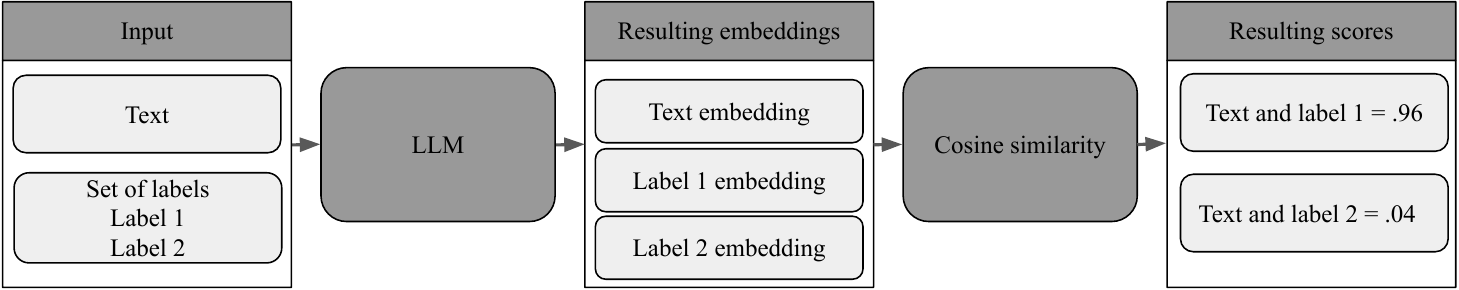}
 \caption{An illustration of embedding-based \ac{ZSL}}
\label{fig:EBZSL}
\end{figure}

Figure \ref{fig:EBZSL} illustrates the process of embedding-based \ac{ZSL} text classification. Both the input text and potential class labels are passed through a pre-trained \ac{LLM} to generate embeddings. Classification is performed by calculating the cosine similarity between the input text embedding and each class label embedding. The class with the highest similarity score is then selected as the predicted label (i.e., label 1 in the example).

\subsection{\Acf{NLI}-based \ac{ZSL}}
\Ac{NLI}-based \ac{ZSL} frames text classification as a textual entailment problem, where it determines whether a given text (i.e., premise) logically follows another hypothesis. The input text is treated as the \ac{NLI} premise, and each candidate label forms a hypothesis. The model calculates probabilities for entailment and contradiction, which are then converted into label probabilities. The text is classified under the label with the highest entailment probability \cite{yin2019benchmarking, alammar2024hands}.

\begin{figure} [htbp]
\center
 \includegraphics[width=.6\textwidth]{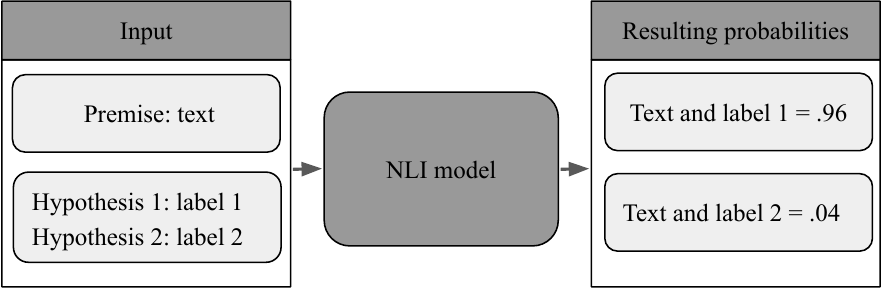}
 \caption{An illustration of \ac{NLI}-based \ac{ZSL}}
\label{fig:NLI}
\end{figure}

As Figure \ref{fig:NLI} presents, the input text serves as the premise, and potential class labels are hypotheses. The \ac{NLI} model assesses whether the premise entails or contradicts each hypothesis and assigns a probability to each case.
The input text is classified based on the label with the highest entailment probability, in this case, label 1.

\subsection{\Acf{TARS}-based \ac{ZSL}}
\ac{TARS} formulates the classification task as a universal binary classification problem, where the model learns to predict whether a given text belongs to a particular label or not. Instead of training separate models for each label, \ac{TARS} simultaneously evaluates the relevance of the text for all labels by adapting \ac{LLM} representations through label-conditioned embeddings \cite{halder2020task}.

\begin{figure} [htbp]
\center
 \includegraphics[width=.75\textwidth]{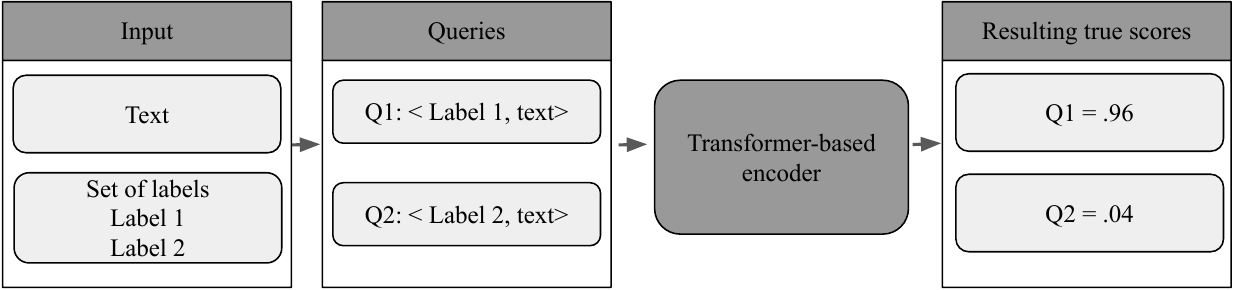}
 \caption{An illustration of \ac{TARS}-based \ac{ZSL}}
\label{fig:TARS}
\end{figure}

As Figure \ref{fig:TARS} illustrates, the input to \ac{TARS} consists of the text to be classified and a set of candidate labels. \ac{TARS} generates embeddings conditioned on both the text and each label by appending the label to the text to form queries. These queries are processed by a shared transformer encoder to produce task-specific embeddings that capture the semantic relationships between the text and the labels. A binary prediction of true or false is then performed for each label. The label with the highest true confidence is selected as the final classification (i.e., label 1 in the example) \cite{halder2020task}.

\subsection{Generative-based \ac{ZSL}}
Transformer-based generative models, such as 
OpenAI 's Generative Pre-Trained Transformers (GPTs) \footnote{https://openai.com/}, are capable to perform \ac{ZSL} text classification by generating text in response to provided input \cite{brown2020language}.

\begin{figure} [htbp]
\center
 \includegraphics[width=.6\textwidth]{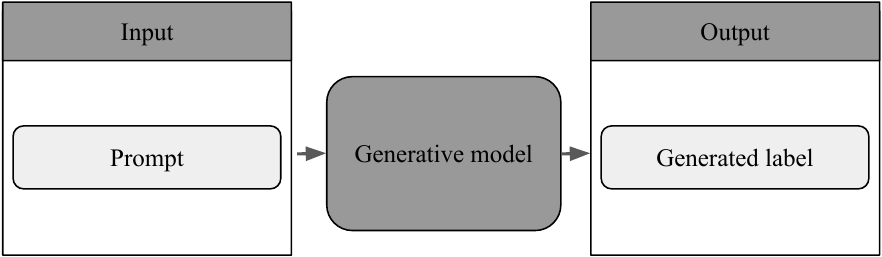}
 \caption{An illustration of generative-based \ac{ZSL}}
\label{fig:GBZSL}
\end{figure}

In this approach, as Figure \ref{fig:GBZSL} depicts, the model receives a prompt that provides specific instructions on how to classify input text and the input text. The provided instruction is in natural language, and it may not include any demonstrations. The model then generates a response that indicates the most appropriate class based on the provided prompt \cite{brown2020language,alammar2024hands}.

\section{Related work} \label{Related work} 
Many studies have assessed sentiment analysis tools, explored the impact of sentiment on software development practices, and more. Due to space constraints, we focus on summarizing (1) systematic reviews related to sentiment analysis tools in software engineering and (2) research efforts on the development of such tools.

\subsection{Systematic reviews on sentiment analysis tools for software engineering}

S{\'a}nchez-Gord{\'o}n and Colomo-Palacios \cite{sanchez2019taking} conducted a \ac{SLR} on software developers' emotions. The study highlighted the limited research in this domain and noted that although current approaches are recognized as unreliable, many techniques with the potential to enhance the detection of developers’ emotions remain underutilized or unexplored.

Obaidi and Kl{\"u}nder \cite{obaidi2021development} conducted an \ac{SLR} on software engineering sentiment analysis tools. The study found that most research relies on using existing tools, \ac{SVM} is the most utilized technique, and \ac{OSS} projects are the main data source. The study highlighted the challenges of training data scarcity, inconsistent tool performance, and the subjectivity of annotated data. The study also noted that sarcasm and irony detection remains a major challenge in sentiment analysis for software engineering.

In a follow-up \ac{SMR}, Obaidi et al. \cite{obaidi2022sentiment} expanded the previous analysis to include recent sentiment analysis studies in software engineering. The study confirmed that research still predominantly applies existing sentiment analysis tools rather than developing new ones, \ac{OSS} data remains the most used soure, \ac{SVM} and \ac{GBT} were identified as the most common supervised learning techniques, and that fine-tuned transformer models outperform other approaches. The study also highlighted  the unreliability of general sentiment analysis tools and the need for customization to specific software engineering contexts.
%%%%%%%%% stopped here: 

Lin et al. \cite{lin2022opinion} conducted an \ac{SLR} that identified sentiment analysis tools currently in use and raised concerns about their application in unintended domains without proper validation. The review summarized comparisons of these tools, identified publicly available datasets, and highlighted challenges in software engineering sentiment analysis, such as identifying neutral sentiment. The study emphasized that model quality depends on training dataset quality and noted the considerable effort needed to train supervised machine learning models.

\subsection{Sentiment analysis tools for software engineering}
%%%Make it short

Several sentiment analysis tools have been developed for software engineering utilizing a variety of \ac{NLP} and machine learning approaches. Heuristic-based methods were employed by Islam et al. \cite{islam2018sentistrength} and Islam and Zibran \cite{islam2018deva}. \ac{SVM} were utilized by Calefato et al. \cite{calefato2018sentiment}, Islam et al. \cite{islam2019marvalous}, Murgia et al. \cite{murgia2018exploratory}, and Cagnoni et al. \cite{cagnoni2020emotion}. Ensemble approaches were adopted by Uddin et al. \cite{uddin2022empirical}, Ahmed et al. \cite{ahmed2017senticr}, and Ding et al. \cite{ding2018entity}.

%Additionally, Chen et al. \cite{chen2019sentimoji} incorporated emojis in sentiment analysis. Opinion mining and summarization tools were tailored to specific software engineering contexts by Uddin et al. \cite{uddin2017opiner}, Lin et al. \cite{lin2019pattern}, and Calefato et al. \cite{calefato2019emtk}.

%%%%%%%%%%%%%%%%%
%%%%%%%%%%%%%%%%%%
%%%%%%%%%%%%%%%%%%%

Aiming to leverage transformer-based models, Biswas et al. \cite{biswas2020achieving} introduced BERT4SentiSE, a BERT-based sentiment classifier fine-tuned on \ac{SO} posts. Zhang et al. \cite{zhang2020sentiment} assessed the performance  of BERT, RoBERTa, XLNet, and ALBERT pre-trained transformer models across various software engineering contexts, and they found that fine-tuning these models outperforms state-of-the-art sentiment analysis tools. Similarly, Batra et al. \cite{batra2021bert} evaluated fine-tuned BERT, ensemble BERT models, and compressed BERT for sentiment analysis on \ac{SO} posts, GitHub commit comments, and Jira issue comments. The study found that compressed BERT and ensemble BERT produced better similar results, with the compressed version recommended for resource conservation. Bleyl et al. \cite{bleyl2022emotion} developed a fine-tuned BERT model for detecting emotions in \ac{SO} posts. Additionally, Sun et al. \cite{sun2022incorporating} introduced EASTER, a sentiment analysis tool that integrates RoBERTa as the embedding layer in TextCNN. EASTER was evaluated on app reviews, Jira issue comments, and \ac{SO} posts.

%%%%%%%%%%%%%%%%%%%%%%%%%%%%%%%%%%%%%%%%%%%
%%%%%%%%%%%%%%%%%%%%%%%%%%%%%%%%%%%%%%%%%%%
%%%%%%%%%%%%%%%%%%%%%%%%%%%%%%%%%%%%%%%%%%%
%%%%%%%%%%%%%%%%%%%%%%%%%%%%%%%%%%%%%%%%%%%
%%%%%%%%%%%%%%%%%%%%%%%%%%%%%%%%%%%%%%%%%%%
Shafikuzzaman et al. \cite{shafikuzzaman2024empirical} evaluated the performance of twelve pretrained language models, including fine-tuned models, SentiStrength-SE, and SentiCR \cite{ahmed2017senticr} on the Gerrit, GitHub, Google Play, Jira, and \ac{SO} posts datasets.
The study found that model performance varied across datasets, with fine-tuned models performing better on larger datasets. To better understand models’ behavior, the study used Shapley Additive Explanations (SHAP) to conduct error analysis. Leveraging generative-based models, Zhang et al. \cite{zhang2023revisiting} explored the potential of Llama 2-Chat; Vicuna; and WizardLM, referred to as \acp{bLLM}, for sentiment analysis. The study assessed their performance using \ac{ZSL} and \ac{FSL} on sentiment analysis tasks across five datasets: Gerrit, GitHub, Google Play, Jira, and \ac{SO}. Due to cost concerns, the study was conducted on a stratified representative sample drawn from 10\% of each dataset that represents the test set. The findings indicated no significant performance difference between \ac{ZSL} and \ac{FSL} and that \ac{FSL} did not necessarily outperform \ac{ZSL}. The analysis was extended by fine-tuning BERT, RoBERTa, ALBERT, XLNet, and DistilBERT, referred to as \acp{sLLM}. The comparison of the \acp{sLLM}
 to the \acp{bLLM} revealed that \acp{bLLM} outperformed \acp{sLLM} on imbalanced datasets or those with limited training data, while fine-tuned \acp{sLLM} performed better when ample training data and balanced distributions were available.

\section{Study setup} \label{Study setup} 
This section describes the setup of the study, including the goal and research questions (RQs), datasets, selected \acp{LLM}, label curation and configuration, performance measures, statistical analysis, and implementation details. 

\subsection{Goal and research questions (RQs)} 
The goal of this study is defined using the Goal-Question-Metric (GQM) template \cite{caldiera1994goal}, as follows: 

Assessing the performance of \ac{ZSL} in sentiment classification within the context of API reviews, code review comments, pull requests and commit comments, developer messages, mobile app reviews, issue comments, and posts on technical question-and-answer websites in the software engineering domain. 
To achieve this goal, we formulated the following RQs: 

\begin{itemize} 
\item \textbf{RQ1:} Which \ac{ZSL} technique is most effective for sentiment classification, and among the techniques that evaluate multiple models, which model demonstrates the best performance?
 
 \item \textbf{RQ2:} Do different label configurations have an impact on the performance of \ac{ZSL}-based sentiment classification? 
 
\item \textbf{RQ3:} How does the performance of \ac{ZSL}-based models compare with that of the state-of-the-art fine-tuned transformer-based models in sentiment classification?

 \item \textbf{RQ4:} What factors contribute to the misclassification of sentiment labels in \ac{ZSL}-based models, and how do these compare with those shared with the state-of-the-art fine-tuned transformer-based models?
\end{itemize}

\subsection{Datasets} \label{Dataset}
To address the RQs, we utilized seven publicly available datasets commonly used for sentiment analysis in software engineering. Table \ref{tab:datasetsummary} summarizes these datasets, and a description of each is provided below.

\textbf{API reviews:} %%% okay
The dataset, curated by Uddin and Khomh \cite{uddin2019automatic}, contains 4,522 sentences from 1,338 \ac{SO} posts across 71 threads tagged with 18 Java API-related keywords. %Seven coders participated in the annotation, with each sentence initially annotated by three. Disagreements were resolved through majority voting or by a fourth coder. 
The  dataset includes 890 positive, 496 negative, and 3,136 neutral sentences.

\textbf{Gerrit:} %%% okay
The dataset, curated by Ahmed et al. \cite{ahmed2017senticr}, includes 1,600 code review comments mined from the code review repositories of 20 open-source projects, with 398 labeled as negative and 1,202 as non-negative. %The comments were . %Three authors independently annotated each comment as positive, neutral, or negative. Inconsistencies were resolved through joint discussions, and positive and neutral comments were later reclassified as non-negative.

\textbf{GitHub:} 
The dataset, curated by Novielli et al. \cite{novielli2020can}, consists of 7,122 GitHub pull request and commit comments, labeled as 2,013 positive, 2,087 negative, and 3,022 neutral. %Initially, 4,000 comments from \cite{pletea2014security} were annotated by two authors, who resolved inconsistencies and discarded 69 comments, leaving 3,931 for retraining the Senti4SD tool \cite{calefato2018sentiment}. The tool then annotated 112,000 comments, with a subset of 4,809 manually validated and merged with the previous set. The tool was subsequently used to annotate another 111,000 comments, with another subset manually verified, resulting in the final dataset of 7,122 comments.

\textbf{Gitter:} % I think we did the mapping?
The dataset, curated by Sajadi et al. \cite{sajadi2023towards}, contains 400 developer messages from 10 Gitter communities. Messages were annotated for six basic emotions (i.e., anger, love, fear, joy, sadness, and surprise) and subcategories based on Shaver’s emotion taxonomy \cite{imran2022data}. 
Following the approach of \cite{lin2018sentiment}, we mapped love and joy to positive, and anger and sadness to negative, resulting in 201 messages with 127 positive and 74 negative.

\textbf{Google Play:} %%%OK 
The dataset, curated by Lin et al. \cite{lin2018sentiment}, contains 341 Android app reviews from Google Play, with 186 labeled as positive, 130 as negative, and 25 as neutral. %It was derived from a larger dataset of 3,000 reviews by \cite{villarroel2016release}, from which \cite{lin2018sentiment} selected a random sample. Two authors annotated each review as positive, neutral, or negative, with inconsistencies resolved by a third author.

\textbf{Jira:} %%%OK
The dataset, curated by Lin et al. \cite{lin2018sentiment}, contains 926 sentences from Jira issue comments, with 636 labeled as negative and 290 as positive. %It is derived from \cite{ortu2016emotional}’s dataset, originally annotated for six emotions: love, joy, surprise, anger, sadness, and fear, where love and joy were mapped as positive and anger and sadness as negative.

\textbf{\ac{SO}:} %%%OK
The dataset, curated by Calefato et al. \cite{calefato2018sentiment}, includes
4,423 \ac{SO} posts, with 1,527 positive, 1,202 negative, and 1,694 neutral.

\begin{table}[htbp]
\centering
\scriptsize
\caption{Summary of utilized datasets}
\label{tab:datasetsummary} 
\begin{tabular}{@{}llll@{}}
\toprule
\textbf{Dataset Name} & \textbf{Total} & \textbf{Polarity distribution} \\ \midrule
API reviews & 4,522 & Positive (890), negative (496), and neutral (3,136) \\
Gerrit & 1,600 & Negative (398) and non-negative (1,202) \\
GitHub & 7,122 & Positive (2,013), negative (2,087), and neutral (3,022) \\
Gitter & 201 & Positive (127) and negative (74) \\
Google Play & 341 & Positive (186), negative (130), and neutral (25) \\
Jira & 926 & Positive (290) and negative (636) \\
\ac{SO} & 4,423 & Positive (1,527), negative (1,202), and neutral (1,694) \\ 
\bottomrule
\end{tabular}
\end{table}

\subsection{\Acf{LLM} selection} 
To address the aforementioned research questions (RQs), we selected the models below. Table \ref{tab:model} summarizes these models, with unique identifier for easy reference throughout the study. 

Our selection was guided by four key considerations. First, we prioritized reproducibility and accessibility by including publicly available and widely adopted pretrained models such as BERT, RoBERTa, and ALBERT, which are well-established benchmarks in \ac{ZSL} and transfer learning research \cite{alhoshan2023zero,zhang2023revisiting}. Second, to ensure architectural diversity, we selected transformer variants trained with distinct pretraining objectives, including masked language modeling, permutation modeling, and next-sentence prediction. Third, we sought domain variation by incorporating both generic and domain-specific models. For instance, BERTOverflow for technical Q\&A, RoBERTa-base-go\_emotions for emotions, and Twitter-RoBERTa for social media sentiment analysis. Finally, to capture a broad availability spectrum, we included both paid and unpaid models.

\begin{itemize}

\item \textbf{Embedding-based \ac{ZSL}}:We selected the following unpaid, generic-embeddings:  BERT-base-uncased \footnote{https://huggingface.co/google-bert/bert-base-uncased}, RoBERTa-base \footnote{https://huggingface.co/FacebookAI/roberta-base}, DistilBERT-base-uncased \footnote{https://huggingface.co/distilbert/distilbert-base-uncased}, ALBERT-base-v2 \footnote{https://huggingface.co/albert/albert-base-v2}, XLNet-base-cased: \footnote{https://huggingface.co/xlnet/xlnet-base-cased}, and All-MiniLM-L12-v2 \footnote{https://huggingface.co/sentence-transformers/all-MiniLM-L12-v2}.

Besides the aforementioned unpaid, generic models, we utilized the following unpaid, domain-specific models:  BERTOverflow \footnote{https://huggingface.co/jeniya/BERTOverflow},RoBERTa-base-go\_emotions \footnote{https://huggingface.co/SamLowe/roberta-base-go\_emotions}, and Twitter-RoBERTa-base-sentiment \footnote{https://huggingface.co/cardiffnlp/twitter-roberta-base-sentiment}.

We also included three paid OpenAI embeddings\footnote{https://openai.com/index/new-embedding-models-and-api-updates/}: Text-embedding-ada-002, Text-embedding-3-small, and Text-embedding-3-large.

\item \textbf{\ac{NLI}-based \ac{ZSL}}: To evaluate the performance of \ac{NLI}-based \ac{ZSL}, we selected the following four models that are specialized for \ac{NLI} tasks: RoBERTa-large-mnli \footnote{https://huggingface.co/FacebookAI/roberta-large-mnli},Cross-encoder/nli-deberta-base \footnote{https://huggingface.co/cross-encoder/nli-deberta-base}, BART-large-mnli \footnote{https://huggingface.co/facebook/bart-large-mnli}, and DeBERTa-v3-large-mnli-fever-anli-ling-wanli \footnote{https://huggingface.co/MoritzLaurer/DeBERTa-v3-large-mnli-fever-anli-ling-wanli}.

\item \textbf{\ac{TARS}}: We used the model implementation from the original paper that introduced the technique \cite{halder2020task}.

\item \textbf{Generative-based \ac{ZSL}}: For generative-based \ac{ZSL}, we used GPT-3.5 Turbo (gpt-3.5-turbo-0125\footnote{https://platform.openai.com/docs/models\#gpt-3-5-turbo}), as it was the most viable option in terms of efficiency and cost at the time of conducting the study (i.e., May 2024).

\end{itemize}

\begin{table}[htbp]

\scriptsize
\centering
\caption{Selected models for each \ac{ZSL} technique}
\label{tab:model}
\begin{tabular}{lll}
\toprule
\textbf{Approach} & \textbf{Model} & \textbf{Identifier} \\ 
\midrule
Embedding-based \ac{ZSL} & BERT-base-uncased & E\_M1 \\ 
 & RoBERTa-base & E\_M2 \\ 
 & DistilBERT-base-uncased & E\_M3 \\ 
 & ALBERT-base-v2 & E\_M4 \\ 
 & XLNet-base-cased & E\_M5 \\ 
 & All-MiniLM-L12-v2 & E\_M6 \\ 
 & BERTOverflow & E\_M7 \\ 
 & RoBERTa-base-go\_emotions & E\_M8 \\ 
 & Twitter-RoBERTa-base-sentiment & E\_M9 \\ 
 & Text-embedding-ada-002 & E\_M10 \\ 
 & Text-embedding-3-small & E\_M11 \\ 
 & Text-embedding-3-large & E\_M12 \\ 
\midrule
\ac{NLI}-based \ac{ZSL} & RoBERTa-large-mnli & N\_M1 \\ 
 & Cross-encoder/nli-deberta-base & N\_M2 \\ 
 & BART-large-mnli & N\_M3 \\ 
 & DeBERTa-v3-large-mnli-fever-anli-ling-wanli & N\_M4 \\ 
\midrule
\ac{TARS}-based \ac{ZSL} & \ac{TARS} & T\_M1 \\ 
\midrule
Generative-based \ac{ZSL} & GPT-3.5 Turbo (gpt-3.5-turbo-0125) & G\_M1 \\ 
\bottomrule
\end{tabular}
\end{table}

\subsection{Label curation and configuration} \label{LCC}

To investigate the impact of label configurations on sentiment analysis, we compare three distinct types: the original dataset, expert-curated, and \ac{LLM}-generated labels. These configurations differ in phrasing, contextual specificity, and descriptive granularity. While the original labels perform well on standard benchmarks, we aim to explore whether enriching them with descriptions and contextual cues can improve model performance in a \ac{ZSL} setting. Specifically, including contextual information about the dataset instance type may help models better interpret the input, while adding sentiment descriptors can enhance the semantic richness of the labels and improve embedding quality. 

We include both expert-curated and \ac{LLM}-generated labels to investigate two approaches to enriching label semantics. Expert-curated labels offer human-level domain insight, with the potential to capture subtle distinctions and contextual relevance that may be overlooked in the original labels. In contrast, \ac{LLM}-generated labels provide a scalable, automated alternative that reflects the model’s own interpretation of sentiment. Comparing both approaches allows us to assess the trade-offs between human judgment and automated label generation and to examine whether either leads to improved performance in \ac{ZSL} settings over the original labels.

Table \ref{tab:label_categories} summarizes these labels with identifiers and examples. Moreover, our online appendix includes the full set of utilized labels \footnote{\url{https://osf.io/gzt9r/?view_only=afd4a24f2d724413aa5423eac0cdcfa6}}.
It is important to note that all alternative labels were derived by mapping to the original label set, with no reannotation of the dataset instances involved.

\begin{itemize}
 \item \textbf{Original labels:} We used the original sentiment labels from the datasets as described in Section \ref{Dataset}.

 \item \textbf{Expert-curated labels:} 
The two authors independently created labels based on their understanding of the sentiment classes and dataset context. These labels were then reviewed and consolidated in a joint meeting, where two types of disagreements emerged. Phrase disagreements, in which one author used ``with'' while the other used ``has'' to describe instances.
After discussion, the authors decided that using ``with'' was more appropriate. The other type is
content disagreements, where one author used the original set of emotions mapped to sentiments, while the other did not. The authors decided to retain these labels, as they could improve understanding of the impact of labels.

\item \textbf{\ac{LLM}-generated labels:} We used ChatGPT-3.5 to generate labels using the following prompt: \textit{``Generate a list of words that best describe positive sentiment.''}. The term ``positive'' was replaced with ``negative'' to generate words for negative sentiment, and the results of both were negated to generate neutral labels. The result formed two label configurations: L6 using ChatGPT-suggested words and L7 using a combination of suggested words and corresponding sentiment classes.
\end{itemize}

\begin{table}[htbp]
\centering
\scriptsize
\caption{ Summary of label configurations}
\label{tab:label_categories}
\begin{tabular}{@{}p{0.1\textwidth}p{0.2\textwidth}p{0.4\textwidth}p{0.25\textwidth}@{}}
\toprule
\textbf{Identifier} & \textbf{Category} & \textbf{Description} & \textbf{Example} \\ \midrule
L1 & Original & Labels used as originally described in each dataset, where emotions were mapped to corresponding sentiments (i.e., joy and love to positive; anger and sadness to negative). & Positive 
\\ \midrule

L2 & Expert-curated & Labels that use the term sentiment to describe the type of instances in each dataset, where neutral is described using negations of both sentiments.
 
%“Labels that categorize instances in each dataset by their sentiment (positive, negative, or neutral), where neutral refers to cases that are neither positive nor negative.”
 & A positive app review \\ \midrule

L3 & Expert-curated & 

Labels describing the type of dataset instances with attached sentiment using ``with” along with the term ``sentiment", where neutral is described using negations of both sentiments.

 & An app review with positive sentiment \\ \midrule

L4 & Expert-curated & Labels describing the type of dataset instances with attached sentiment and associated original emotions mapped to the sentiment (i.e., joy and love for positive; anger and sadness for negative) using ``with” and the term ``sentiment" , where neutral is described using negations of both sentiments and their associated emotions. & An app review with positive, joy, or love sentiments

\\ \midrule

L5 & Expert-curated & Labels describing the type of dataset instances with only the emotions mapped to the sentiment (i.e., joy and love for positive; anger and sadness for negative) using ``with”, where neutral is described by negating both the sentiments associated emotions. & An app review with joy or love sentiments \\ \midrule

L6 & \ac{LLM}-generated & Labels describing the type of dataset instances with only words generated by the \ac{LLM} using ``with”, where neutral is described by negating both sentiments associated words. & An app review with cheerfulness, happiness, amusement, satisfaction, bliss, gaiety, glee, jolliness, joviality, joy, delight, enjoyment, gladness, jubilation, elation, ecstasy, euphoria, zest, enthusiasm, excitement, thrill, zeal, exhilaration, contentment, pleasure, and optimism sentiments \\ \midrule

L7 & \ac{LLM}-generated & Labels describing the type of dataset instances with sentiment and words generated by the \ac{LLM} using ``with”, where neutral is described by negating both the sentiments and their associated words. & An app review with positive, cheerfulness, happiness, amusement, satisfaction, bliss, gaiety, glee, jolliness, joviality, joy, delight, enjoyment, gladness, jubilation, elation, ecstasy, euphoria, zest, enthusiasm, excitement, thrill, zeal, exhilaration, contentment, pleasure, and optimism sentiments \\

\\ \bottomrule
\end{tabular}
\end{table}

\subsection{Performance measures}
To evaluate the performance of the models, we calculated both macro-F1 and micro-F1 scores, which are variations of the F1 score \cite{manning2008introduction}, following the approach of previous, related work \cite{novielli2020can,zhang2023revisiting}. Macro-F1 calculates the F1 score for each class independently and averages them, while micro-F1 aggregates the contributions of all classes to compute the average.

\subsection{Statistical analysis}
Merely comparing performance measures is insufficient, as observed differences may arise due to random variability \cite{witten2005practical}. To assess the significance of these differences, we use the non-parametric Scott-Knott Effect Size Difference (ESD) test, which produces distinct, non-overlapping groups and quantifies the magnitude of meaningful median differences \cite{tantithamthavorn2018impact}. The test is robust against outliers and does not assume homogeneity, normality, or sample size \cite{puth2015effective}, and it has been successfully applied in similar contexts \cite{tantithamthavorn2018impact}.

While we report both macro-F1 and micro-F1 scores for comprehensive evaluation, we base our comparisons on the macro-F1 score, as it better handles imbalanced datasets by giving equal weight to all classes, aligning with previous studies \cite{manning2008introduction, novielli2020can, zhang2023revisiting}.

\subsection{Implementation} 

To implement the empirical assessment, we followed a series of steps for each RQ.

For RQ1 and RQ2, we used all datasets, models, and label configurations as described above. For the generative-based \ac{ZSL}, we used the template:
\textit{``What is the sentiment of the following app review, which is delimited with triple backticks?'' Give your answer as either `positive', `negative', or `neutral'.''}
We replaced the term ``app review" with the type of each examined dataset and substituted the labels ``positive," ``negative," or ``neutral" with the specific labels being assessed.
We used the default parameters provided by the generative model’s API, with the exception of setting the temperature to zero to reduce variability in the outputs.

To ensure consistency between the model outputs and the gold labels, we applied simple post-processing rules. For the shorter labels, we checked whether the generated output explicitly mentioned the sentiment name (i.e., positive, negative, neutral) and mapped it to the corresponding label. For the longer labels (i.e., L6 and L7), the model occasionally produced partial matches by omitting certain parts; in these cases, we post-processed the outputs to align them with the original labels.

For RQ3, following \cite{zhang2023revisiting}, we partitioned each dataset into training, validation, and test sets  in an 8:1:1 ratio with stratified splitting. We fine-tuned state-of-the-art transformer models listed in Table \ref{tab:RQ3-Set}.

We note that our evaluation does not include earlier SE-specific sentiment analysis tools, such as SentiStrength-SE. This decision is supported by the findings of Zhang et al. \cite{zhang2020sentiment}, who demonstrated that fine-tuned transformer-based models outperform these tools and concluded that such models should be regarded as the state of the art for sentiment analysis in the SE domain. Furthermore, in a more recent work \cite{zhang2023revisiting}, Zhang et al. followed this conclusion by excluding these earlier tools entirely and focusing their evaluation solely on fine-tuned transformer-based models. In line with this direction, we adopt a similar approach and benchmark our results against state-of-the-art transformer models.

 We used a learning rate of $2 \times 10^{-5}$, 5 epochs, batch size of 32, and a max sequence length of 256 tokens. The model with the highest macro-F1 score on the validation set was evaluated on the test set with original labels (i.e., L1). We compared the performance of fine-tuned models with the best-performing \ac{ZSL} model-label combinations and the best-performing model for each \ac{ZSL} technique when paired with L1.

For RQ4, we conducted quantitative and qualitative analyses on misclassifications from RQ3. The quantitative analysis identified common misclassified instances among \ac{ZSL}-based and fine-tuned models. The qualitative analysis categorized these misclassifications using the framework of Novielli et al. \cite{novielli2018benchmark}. We independently categorized the commonly misclassified instances of the \ac{ZSL}-based models. Then we compared the results in a joint session and calculated Cohen's kappa coefficient, which was 0.71, indicating moderate agreement \cite{mchugh2012interrater}. Disagreements were resolved through discussion, with each author providing justification until consensus was reached.

% Please add the following required packages to your document preamble:
% \usepackage{booktabs}
\begin{table}[htbp]
\scriptsize
\caption{Selected models for fine-tuning}
\label{tab:RQ3-Set}
\begin{tabular}{@{}ll@{}}
\toprule
\textbf{Model} & \textbf{Identifier} \\ \midrule
BERT-base-cased & F\_M1 \\
RoBERTa-base & F\_M2 \\
DistilBERT-base-uncased & F\_M3 \\
ALBERT-base-v1 & F\_M4 \\
XLNet-base-cased & F\_M5 \\
\bottomrule
\end{tabular}
\end{table}

\section{Results}\label{Results} 
This section summarizes the results of the study, and our online appendix includes more detailed results for each RQ. 

\subsection{RQ1: Which \ac{ZSL} technique is most effective for sentiment classification, and among the techniques that evaluate multiple models, which model demonstrates the best performance?}
The results of the examined \ac{ZSL} techniques on each dataset are summarized in Table \ref{tab:my-table}. The ``Mac" and ``Mic" columns refer to the macro-F1 score and micro-F1 score values, respectively, with the highest values of the macro-F1 and micro-F1 scores for each dataset are bolded. 

As presented in the table, generative-based \ac{ZSL} achieved the highest macro-F1 scores across most datasets, with exceptions in the Gitter, Google Play, and Jira datasets. N\_M4 and N\_M2 outperformed others on the Gitter and Google Play datasets, respectively, and E\_M9 achieved the highest macro-F1 score for the Jira dataset.

Within the embedding-based \ac{ZSL} models, E\_M9 achieved the highest macro-F1 scores across most datasets (i.e., 5 out of 7). Among the \ac{NLI}-based models, N\_M2 can be considered the best performer, where it achieved the highest macro-F1 score in 5 out of 7 datasets.

\begin{figure} [hbtp]
 \center
 \includegraphics[width=.98
 \textwidth]{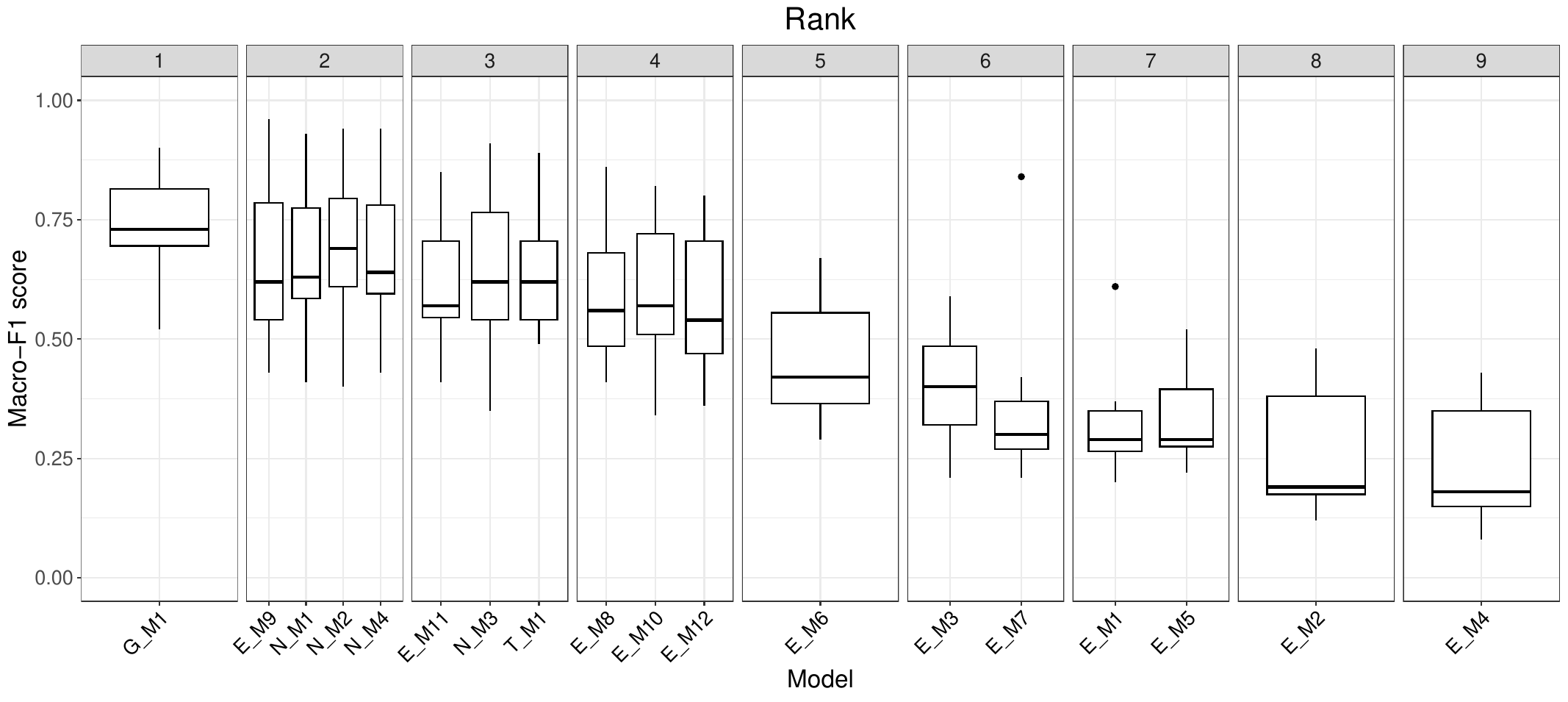}
 \caption{Scott-Knott ESD ranking for \ac{ZSL} models based on macro-F1 score}
 \label{fig:RQ1}
\end{figure}

The results of the statistical test confirmed the above findings, where Figure \ref{fig:RQ1} depicts the results. The generative-based \ac{ZSL} model was ranked as the best performing model among all examined models. The second rank predominantly comprised \ac{NLI}-based \ac{ZSL} models and one embedding-based \ac{ZSL} model (i.e., E\_M9).
The third rank included the remaining \ac{NLI}-based \ac{ZSL} model (i.e., N\_M3), the \ac{TARS}-based model, and another embedding-based model (i.e., E\_M11). The rest of the embedding-based models were distributed from the fourth rank to the last rank (i.e., rank 9).

The results highlight the dominance of generative-based and \ac{NLI}-based \ac{ZSL} techniques, with a few embedding-based models demonstrating competitive performance.

\begin{table}[htbp]
\centering
\scriptsize
\begin{tabular}{@{}lcccccccccccccc@{}}
\toprule
\textbf{Model} & \multicolumn{2}{c}{\textbf{API}} & \multicolumn{2}{c}{\textbf{Gerrit}} & \multicolumn{2}{c}{\textbf{GitHub}} & \multicolumn{2}{c}{\textbf{Gitter}} & \multicolumn{2}{c}{\textbf{Google}} & \multicolumn{2}{c}{\textbf{Jira}} & \multicolumn{2}{c}{\textbf{\ac{SO}}} \\ 
 & \multicolumn{2}{c}{\textbf{reviews}} & \multicolumn{2}{c}{} & \multicolumn{2}{c}{} & \multicolumn{2}{c}{} & \multicolumn{2}{c}{\textbf{Play}} & \multicolumn{2}{c}{} & \multicolumn{2}{c}{\textbf{}} \\ 
 \midrule
 & \textbf{Mac} & \textbf{Mic} & \textbf{Mac} & \textbf{Mic} & \textbf{Mac} & \textbf{Mic} & \textbf{Mac} & \textbf{Mic} & \textbf{Mac} & \textbf{Mic} & \textbf{Mac} & \textbf{Mic} & \textbf{Mac} & \textbf{Mic} \\ 
 \midrule
\multicolumn{15}{c}{Embedding-based \ac{ZSL}} \\ \midrule
E\_M1 & 0.37 & 0.48 & 0.2 & 0.25 & 0.26 & 0.39 & 0.29 & 0.38 & 0.27 & 0.36 & 0.61 & 0.75 & 0.33 & 0.4 \\
E\_M2 & 0.12 & 0.13 & 0.43 & 0.75 & 0.18 & 0.3 & 0.33 & 0.39 & 0.19 & 0.38 & 0.48 & 0.7 & 0.17 & 0.28 \\
E\_M3 & 0.37 & 0.48 & 0.21 & 0.26 & 0.4 & 0.41 & 0.47 & 0.52 & 0.27 & 0.36 & 0.59 & 0.6 & 0.5 & 0.5 \\
E\_M4 & 0.08 & 0.12 & 0.43 & 0.75 & 0.16 & 0.3 & 0.29 & 0.38 & 0.18 & 0.38 & 0.41 & 0.69 & 0.14 & 0.27 \\
E\_M5 & 0.22 & 0.24 & 0.5 & 0.62 & 0.29 & 0.31 & 0.29 & 0.38 & 0.27 & 0.36 & 0.52 & 0.58 & 0.28 & 0.33 \\
E\_M6 & 0.29 & 0.34 & 0.49 & 0.53 & 0.42 & 0.42 & 0.62 & 0.62 & 0.38 & 0.48 & 0.67 & 0.73 & 0.35 & 0.36 \\
E\_M7 & 0.26 & 0.33 & 0.21 & 0.25 & 0.32 & 0.36 & 0.84 & 0.86 & 0.28 & 0.44 & 0.42 & 0.42 & 0.3 & 0.38 \\
E\_M8 & 0.41 & 0.49 & 0.43 & 0.63 & 0.56 & 0.57 & 0.86 & 0.86 & 0.54 & 0.69 & 0.7 & 0.75 & 0.66 & 0.67 \\
E\_M9 & 0.48 & 0.5 & 0.43 & 0.75 & 0.6 & 0.61 & 0.9 & 0.9 & 0.62 & 0.87 & \textbf{0.96} & \textbf{0.96} & 0.67 & 0.69 \\
E\_M10 & 0.34 & 0.35 & 0.5 & 0.51 & 0.52 & 0.52 & 0.82 & 0.84 & 0.63 & 0.75 & 0.81 & 0.82 & 0.57 & 0.58 \\
E\_M11 & 0.41 & 0.49 & 0.55 & 0.72 & 0.54 & 0.54 & 0.84 & 0.85 & 0.57 & 0.7 & 0.85 & 0.86 & 0.57 & 0.57 \\
E\_M12 & 0.36 & 0.4 & 0.46 & 0.47 & 0.48 & 0.49 & 0.79 & 0.81 & 0.62 & 0.77 & 0.8 & 0.81 & 0.54 & 0.56 \\
 \midrule
\multicolumn{15}{c}{\ac{NLI}-based \ac{ZSL}} \\ \midrule
N\_M1 & 0.41 & 0.42 & 0.66 & 0.71 & 0.56 & 0.57 & 0.89 & 0.9 & 0.61 & 0.85 & 0.93 & 0.94 & 0.63 & 0.65 \\
N\_M2 & 0.4 & 0.43 & 0.69 & 0.76 & 0.58 & 0.58 & 0.88 & 0.88 & \textbf{0.71} & 0.87 & 0.94 & 0.95 & 0.64 & 0.67 \\
N\_M3 & 0.35 & 0.35 & 0.62 & 0.7 & 0.5 & 0.53 & 0.9 & 0.9 & 0.63 & 0.88 & 0.91 & 0.92 & 0.58 & 0.62 \\
N\_M4 & 0.43 & 0.46 & 0.65 & 0.69 & 0.56 & 0.59 &\textbf{ 0.91} & \textbf{0.91} & 0.64 & \textbf{0.89} & 0.94 & 0.94 & 0.63 & 0.66 \\
\midrule
\multicolumn{15}{c}{\ac{TARS}-based \ac{ZSL}} \\ \midrule
T\_M1 & 0.49 & \textbf{0.63} & 0.64 & 0.7 & 0.49 & 0.52 & 0.77 & 0.78 & 0.59 & 0.67 & 0.89 & 0.9 & 0.62 & 0.62 \\
\midrule
\multicolumn{15}{c}{Generative-based \ac{ZSL}} \\ \midrule
G\_M1 & \textbf{0.52} & 0.58 & \textbf{0.75} & \textbf{0.8} & \textbf{0.73} & \textbf{0.73} & 0.9 & \textbf{0.91} & 0.67 & \textbf{0.89} & 0.88 & 0.89 & \textbf{0.72} & \textbf{0.73}
\\
\bottomrule
\caption{Summary of the performance of \ac{ZSL}-based models}
\label{tab:my-table}
\end{tabular}
\end{table}

\subsection{RQ2: Do different label configurations have an impact on the performance of \ac{ZSL}-based sentiment classification? }

The results of varying the utilized labels within each embedding-based model are presented in Table \ref{tabel:RQ2EL}, where the highest values for both macro-F1 and micro-F1 scores for each dataset are bolded. We note that, due to space limitations, the table only includes model-label combinations that yielded the highest macro-F1 score for a dataset within each model. We provide the results of all model-label combinations in our online appendix.

%%TODO:do we need this?? When analyzing how often each label achieved the highest macro-F1 score when paired with an embedding-based model on each dataset, L3 combinations emerged as the most frequent top performer, achieving the highest macro-F1 score values in 31 instances across all datasets. L2 achieved this in 25 instances, L4 in 16 instances, and both L1 and L5 in 8 instances each. L6 and L7 produced the highest macro-F1 score values in 5 and 4 instances, respectively. 

Two label combinations of E\_M9 achieved the highest macro-F1 scores in the majority of datasets, with the exception of Google Play, where the combination E\_M12\_L3 attained the highest macro-F1 score. Specifically, E\_M9\_L1 achieved the highest value on the Jira dataset, while E\_M9\_L3 achieved the highest macro-F1 scores across the remaining five datasets.

\begin{figure} [hbtp]
 \center
 \includegraphics[width=.98
 \textwidth]{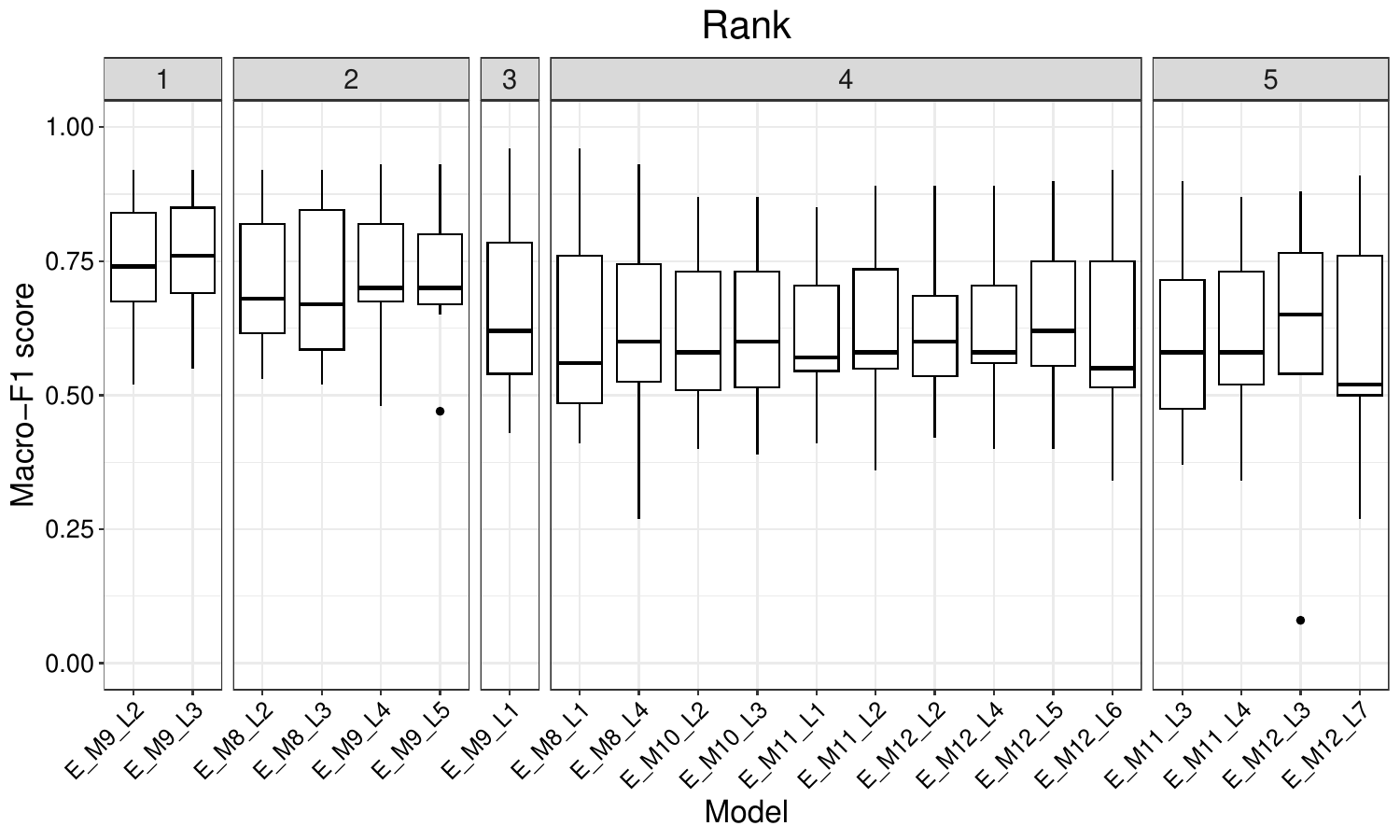}
 \caption{Scott-Knott ESD ranking for embedding-based model-label combinations based on macro-F1 score}
 \label{fig:RQ2_E}
\end{figure}

\begin{table}[htbp]
\centering
\scriptsize

\begin{tabular}{@{}lcccccccccccccc@{}}
\toprule
\textbf{M\&L} & \multicolumn{2}{c}{\textbf{API}} & \multicolumn{2}{c}{\textbf{Gerrit}} & \multicolumn{2}{c}{\textbf{GitHub}} & \multicolumn{2}{c}{\textbf{Gitter}} & \multicolumn{2}{c}{\textbf{Google}} & \multicolumn{2}{c}{\textbf{Jira}} & \multicolumn{2}{c}{\textbf{\ac{SO}}} \\ 
 & \multicolumn{2}{c}{\textbf{reviews}} & \multicolumn{2}{c}{} & \multicolumn{2}{c}{} & \multicolumn{2}{c}{} & \multicolumn{2}{c}{\textbf{Play}} & \multicolumn{2}{c}{} & \multicolumn{2}{c}{\textbf{}} \\ 
 \midrule
 & \textbf{Mac} & \textbf{Mic} & \textbf{Mac} & \textbf{Mic} & \textbf{Mac} & \textbf{Mic} & \textbf{Mac} & \textbf{Mic} & \textbf{Mac} & \textbf{Mic} & \textbf{Mac} & \textbf{Mic} & \textbf{Mac} & \textbf{Mic} \\ 
 \midrule

E\_M1\_L2 & 0.22 & 0.25 & 0.2 & 0.25 & 0.23 & 0.44 & 0.54 & 0.55 & 0.05 & 0.07 & 0.72 & 0.72 & 0.28 & 0.38 \\
E\_M1\_L3 & 0.45 & 0.57 & 0.2 & 0.25 & 0.49 & 0.5 & 0.43 & 0.46 & 0.04 & 0.07 & 0.9 & 0.92 & 0.48 & 0.53 \\
E\_M1\_L4 & 0.22 & 0.23 & 0.43 & 0.75 & 0.27 & 0.33 & 0.29 & 0.38 & 0.37 & 0.47 & 0.59 & 0.74 & 0.32 & 0.33 \\
E\_M1\_L7 & 0.16 & 0.21 & 0.45 & 0.71 & 0.15 & 0.28 & 0.39 & 0.63 & 0.27 & 0.55 & 0.24 & 0.31 & 0.27 & 0.39 \\
 \midrule
E\_M2\_L2 & 0.34 & 0.43 & 0.45 & 0.68 & 0.34 & 0.37 & 0.62 & 0.71 & 0.28 & 0.52 & 0.45 & 0.49 & 0.33 & 0.37 \\
E\_M2\_L3 & 0.16 & 0.19 & 0.44 & 0.74 & 0.44 & 0.51 & 0.65 & 0.68 & 0.38 & 0.47 & 0.43 & 0.68 & 0.32 & 0.36 \\
E\_M2\_L4 & 0.2 & 0.22 & 0.44 & 0.55 & 0.4 & 0.46 & 0.63 & 0.64 & 0.29 & 0.38 & 0.65 & 0.66 & 0.35 & 0.39 \\
E\_M2\_L5 & 0.28 & 0.4 & 0.47 & 0.72 & 0.35 & 0.46 & 0.42 & 0.43 & 0.24 & 0.29 & 0.64 & 0.68 & 0.28 & 0.37 \\
 \midrule
E\_M3\_L1 & 0.37 & 0.48 & 0.21 & 0.26 & 0.4 & 0.41 & 0.47 & 0.52 & 0.27 & 0.36 & 0.59 & 0.6 & 0.5 & 0.5 \\
E\_M3\_L2 & 0.27 & 0.26 & 0.43 & 0.75 & 0.59 & 0.59 & 0.56 & 0.57 & 0.05 & 0.07 & 0.74 & 0.75 & 0.56 & 0.56 \\
E\_M3\_L5 & 0.25 & 0.24 & 0.44 & 0.68 & 0.26 & 0.34 & 0.73 & 0.75 & 0.52 & 0.67 & 0.38 & 0.41 & 0.37 & 0.42 \\
 \midrule
E\_M4\_L2 & 0.37 & 0.48 & 0.44 & 0.72 & 0.6 & 0.62 & 0.86 & 0.86 & 0.16 & 0.16 & 0.38 & 0.38 & 0.44 & 0.44 \\
E\_M4\_L3 & 0.19 & 0.18 & 0.28 & 0.28 & 0.17 & 0.3 & 0.28 & 0.37 & 0.06 & 0.09 & 0.41 & 0.69 & 0.48 & 0.53 \\
E\_M4\_L4 & 0.22 & 0.21 & 0.2 & 0.25 & 0.28 & 0.34 & 0.29 & 0.38 & 0.58 & 0.72 & 0.66 & 0.67 & 0.32 & 0.33 \\

E\_M4\_L5 & 0.09	& 0.12	
&0.2	&0.25 &
0.31	&0.34	

& 0.61	& 0.62	&

0.56	& 0.7	
& 0.37 &	0.4	&
0.29	& 0.32	

\\

 \midrule
E\_M5\_L2 & 0.32 & 0.65 & 0.29 & 0.3 & 0.27 & 0.31 & 0.34 & 0.37 & 0.33 & 0.47 & 0.56 & 0.6 & 0.35 & 0.39 \\
E\_M5\_L3 & 0.35 & 0.45 & 0.36 & 0.36 & 0.4 & 0.41 & 0.55 & 0.56 & 0.29 & 0.35 & 0.53 & 0.61 & 0.38 & 0.38 \\
E\_M5\_L5 & 0.23 & 0.22 & 0.24 & 0.27 & 0.25 & 0.27 & 0.47 & 0.47 & 0.33 & 0.38 & 0.44 & 0.68 & 0.25 & 0.28 \\
E\_M5\_L6 & 0.34 & 0.56 & 0.51 & 0.69 & 0.27 & 0.4 & 0.44 & 0.45 & 0.18 & 0.17 & 0.49 & 0.55 & 0.3 & 0.39 \\

E\_M5\_L7 & 0.34	& 0.55	&0.49	&0.7&0.28	&0.39& 0.38	&0.62	&0.18	&0.18&0.48&	0.55	&0.31	&0.38	\\ \midrule

E\_M6\_L3 & 0.36 & 0.39 & 0.5 & 0.56 & 0.53 & 0.53 & 0.74 & 0.77 & 0.38 & 0.56 & 0.74 & 0.75 & 0.44 & 0.44 \\
E\_M6\_L4 & 0.39 & 0.48 & 0.47 & 0.69 & 0.43 & 0.44 & 0.76 & 0.78 & 0.31 & 0.35 & 0.73 & 0.74 & 0.42 & 0.44 \\
E\_M6\_L7 & 0.33 & 0.54 & 0.46 & 0.5 & 0.46 & 0.46 & 0.72 & 0.74 & 0.51 & 0.68 & 0.8 & 0.82 & 0.43 & 0.46 \\
 \midrule
E\_M7\_L1 & 0.26 & 0.33 & 0.21 & 0.25 & 0.32 & 0.36 & 0.84 & 0.86 & 0.28 & 0.44 & 0.42 & 0.42 & 0.3 & 0.38 \\
E\_M7\_L2 & 0.17 & 0.24 & 0.48 & 0.69 & 0.2 & 0.3 & 0.55 & 0.66 & 0.32 & 0.46 & 0.44 & 0.44 & 0.27 & 0.35 \\
E\_M7\_L3 & 0.22 & 0.28 & 0.43 & 0.75 & 0.27 & 0.44 & 0.58 & 0.65 & 0.23 & 0.36 & 0.47 & 0.47 & 0.31 & 0.39 \\
E\_M7\_L4 & 0.27 & 0.29 & 0.48 & 0.64 & 0.32 & 0.42 & 0.43 & 0.43 & 0.14 & 0.15 & 0.42 & 0.47 & 0.29 & 0.31 \\
E\_M7\_L5 & 0.28 & 0.44 & 0.46 & 0.64 & 0.31 & 0.43 & 0.46 & 0.46 & 0.15 & 0.15 & 0.46 & 0.47 & 0.27 & 0.34 \\
 \midrule

E\_M8\_L2 & 0.53 & 0.65 & 0.61 & 0.77 & 0.68 & 0.69 & 0.9 & 0.91 & 0.62 & 0.75 & 0.7 & 0.74 & 0.74 & 0.75 \\
E\_M8\_L3 & 0.52 & \textbf{0.69} & 0.56 & 0.78 & 0.67 & 0.68 & \textbf{0.91} & 0.91 & 0.61 & 0.74 & 0.86 & 0.87 & 0.78 & \textbf{0.79} \\
 \midrule
E\_M9\_L1 & 0.48 & 0.5 & 0.43 & 0.75 & 0.6 & 0.61 & 0.9 & 0.9 & 0.62 & \textbf{0.87} & \textbf{0.96} & \textbf{0.96} & 0.67 & 0.69 \\

E\_M9\_L2 & 0.52 & 0.65 & \textbf{0.74} & 0.79 & 0.72 & 0.71 & \textbf{0.91} & \textbf{0.92} & 0.63 & 0.85 & 0.92 & 0.93 & 0.77 & 0.77 \\

E\_M9\_L3 & \textbf{0.55} & 0.66 & \textbf{0.74} & \textbf{0.82} & \textbf{0.76} & \textbf{0.76} & \textbf{0.91} & \textbf{0.92} & 0.64 & 0.85 & 0.92 & 0.93 & \textbf{0.79} & 0.79 \\

E\_M9\_L4 & 0.48 & 0.54 & 0.67 & 0.8 & 0.7 & 0.7 & \textbf{0.91} & \textbf{0.92} & 0.68 & 0.79 & 0.93 & 0.93 & 0.73 & 0.73 \\
E\_M9\_L6 & 0.24 & 0.24 & 0.43 & 0.75 & 0.45 & 0.53 & \textbf{0.91} & \textbf{0.92} & 0.64 & 0.86 & 0.92 & 0.93 & 0.48 & 0.59 \\
E\_M9\_L7 & 0.24 & 0.24 & 0.44 & 0.75 & 0.45 & 0.53 & \textbf{0.91} & \textbf{0.92} & 0.62 & 0.86 & 0.92 & 0.93 & 0.48 & 0.59 \\
 \midrule
E\_M10\_L1 & 0.34 & 0.35 & 0.5 & 0.51 & 0.52 & 0.52 & 0.82 & 0.84 & 0.63 & 0.75 & 0.81 & 0.82 & 0.57 & 0.58 \\
E\_M10\_L2 & 0.4 & 0.43 & 0.5 & 0.52 & 0.64 & 0.64 & 0.87 & 0.89 & 0.58 & 0.74 & 0.82 & 0.83 & 0.52 & 0.55 \\
E\_M10\_L3 & 0.39 & 0.41 & 0.57 & 0.6 & 0.46 & 0.51 & 0.87 & 0.88 & 0.61 & 0.79 & 0.85 & 0.86 & 0.6 & 0.61 \\
E\_M10\_L4 & 0.33 & 0.33 & 0.51 & 0.53 & 0.45 & 0.51 & 0.89 & 0.89 & 0.6 & 0.82 & 0.89 & 0.9 & 0.49 & 0.57 \\
E\_M10\_L5 & 0.26 & 0.25 & 0.42 & 0.43 & 0.46 & 0.51 & 0.83 & 0.83 & 0.61 & 0.85 & 0.93 & 0.93 & 0.46 & 0.53 \\
E\_M10\_L6 & 0.3 & 0.29 & 0.48 & 0.54 & 0.49 & 0.5 & 0.89 & 0.9 & 0.63 & 0.81 & 0.72 & 0.73 & 0.48 & 0.53 \\
 \midrule
E\_M11\_L1 & 0.41 & 0.49 & 0.55 & 0.72 & 0.54 & 0.54 & 0.84 & 0.85 & 0.57 & 0.7 & 0.85 & 0.86 & 0.57 & 0.57 \\
E\_M11\_L2 & 0.36 & 0.39 & 0.52 & 0.72 & 0.58 & 0.59 & 0.89 & 0.9 & 0.64 & 0.81 & 0.83 & 0.84 & 0.58 & 0.6 \\
E\_M11\_L3 & 0.37 & 0.42 & 0.44 & 0.45 & 0.51 & 0.52 & 0.9 & 0.91 & 0.65 & 0.86 & 0.78 & 0.78 & 0.58 & 0.6 \\
E\_M11\_L4 & 0.34 & 0.36 & 0.49 & 0.54 & 0.55 & 0.56 & 0.85 & 0.86 & 0.61 & 0.81 & 0.87 & 0.88 & 0.58 & 0.6 \\
 \midrule
E\_M12\_L2 & 0.42 & 0.49 & 0.52 & 0.72 & 0.6 & 0.6 & 0.89 & 0.9 & 0.62 & 0.84 & 0.75 & 0.75 & 0.55 & 0.57 \\
E\_M12\_L3 & 0.08 & 0.12 & 0.54 & 0.55 & 0.54 & 0.56 & 0.88 & 0.89 &\textbf{ 0.69 } & 0.84 & 0.84 & 0.85 & 0.65 & 0.66 \\

E\_M12\_L4 & 0.4 & 0.45 & 0.57 & 0.66 & 0.58 & 0.58 & 0.89 & 0.9 & 0.6 & 0.77 & 0.81 & 0.82 & 0.55 & 0.58 \\
E\_M12\_L5 & 0.4 & 0.54 & 0.54 & 0.59 & 0.63 & 0.63 & 0.87 & 0.88 & 0.62 & 0.82 & 0.9 & 0.91 & 0.57 & 0.61 \\
E\_M12\_L6 & 0.34 & 0.36 & 0.48 & 0.54 & 0.55 & 0.57 & 0.88 & 0.88 & 0.62 & 0.82 & 0.92 & 0.93 & 0.55 & 0.58 

 \\ \bottomrule
 \end{tabular}
\caption{Summary of the performance of best-performing model-label combinations of each embedding-based model across each dataset}
\label{tabel:RQ2EL}
\end{table}
To identify the overall best embedding-based model-label combination, we conducted a statistical analysis. Figure \ref{fig:RQ2_E} presents the results, where only the top five ranked combinations are included due to space constraints. As the figure demonstrates, the analysis confirms the superiority of E\_M9\_L3, as it was ranked as the top-preformer along with E\_M9\_L2. Additionally, E\_M8\_L2, E\_M8\_L3, E\_M9\_L4, and E\_M9\_L5 were ranked in second place while E\_M9\_L1 was ranked third. 
Other combinations followed in the remaining ranks. The analysis revealed that both E\_M8 and E\_M9 achieved higher results with varying label combinations compared to others. 

%%%%NLI
Table \ref{tab:RQ2N} summarizes the results of combining each \ac{NLI}-based model with label configurations, where the highest macro-F1 and micro-F1 scores for each dataset are bolded. N\_M2\_L1 has contributed to the highest macro-F1 score across all \ac{NLI}-model-label combinations in 3 out of 7 datasets. Moreover, N\_M4\_L3 achieved the highest values in 2 datasets. The combinations of N\_M3\_L2, N\_M3\_L3, N\_M3\_L5, N\_M4\_L2, and N\_M4\_L5 were able to achieve the highest macro-F1 score on a dataset. 

%The analysis of how frequently each label achieved the highest macro-F1 score when paired with an \ac{NLI}-based model revealed that L2 combinations were the most top performers, achieving the highest macro-F1 scores in 14 instances across datasets. L3 followed with 9 instances, L1 with 8 instances, and L5 with 3 instances. The remaining labels did not achieve the highest macro value for any dataset when combined with any \ac{NLI}-based model.

\begin{table}[htbp]
\centering
\scriptsize
\begin{tabular}{@{}lcccccccccccccc@{}}
\toprule
\textbf{M\&L} & \multicolumn{2}{c}{\textbf{API}} & \multicolumn{2}{c}{\textbf{Gerrit}} & \multicolumn{2}{c}{\textbf{GitHub}} & \multicolumn{2}{c}{\textbf{Gitter}} & \multicolumn{2}{c}{\textbf{Google}} & \multicolumn{2}{c}{\textbf{Jira}} & \multicolumn{2}{c}{\textbf{\ac{SO}}} \\ 
 & \multicolumn{2}{c}{\textbf{reviews}} & \multicolumn{2}{c}{} & \multicolumn{2}{c}{} & \multicolumn{2}{c}{} & \multicolumn{2}{c}{\textbf{Play}} & \multicolumn{2}{c}{} & \multicolumn{2}{c}{\textbf{}} \\ 
 \midrule
 & \textbf{Mac} & \textbf{Mic} & \textbf{Mac} & \textbf{Mic} & \textbf{Mac} & \textbf{Mic} & \textbf{Mac} & \textbf{Mic} & \textbf{Mac} & \textbf{Mic} & \textbf{Mac} & \textbf{Mic} & \textbf{Mac} & \textbf{Mic} \\ 
 \midrule

N\_M1\_L1 & 0.41 & 0.42 & 0.66 & 0.71 & 0.56 & 0.57 & 0.89 & 0.9 & 0.61 & 0.85 & 0.93 & 0.94 & 0.63 & 0.65 \\
N\_M1\_L2 & 0.4 & 0.41 & 0.68 & 0.74 & 0.58 & 0.59 & 0.92 & \textbf{0.93} & 0.68 & 0.83 & 0.92 & 0.93 & 0.68 & 0.69 \\
N\_M1\_L3 & 0.47 & 0.61 & 0.49 & 0.49 & 0.47 & 0.53 & 0.92 & 0.92 & 0.66 & 0.83 & 0.9 & 0.91 & 0.73 & 0.74 \\
N\_M1\_L4 & 0.22 & 0.22 & 0.27 & 0.3 & 0.46 & 0.52 & 0.89 & 0.9 & 0.59 & 0.82 & 0.9 & 0.91 & 0.47 & 0.56 \\
N\_M1\_L5 & 0.22 & 0.23 & 0.26 & 0.29 & 0.47 & 0.52 & 0.88 & 0.89 & 0.62 & 0.83 & 0.92 & 0.93 & 0.48 & 0.57 \\
N\_M1\_L6 & 0.09 & 0.12 & 0.2 & 0.25 & 0.3 & 0.37 & 0.69 & 0.69 & 0.57 & 0.79 & 0.76 & 0.83 & 0.32 & 0.38 \\
N\_M1\_L7 & 0.11 & 0.13 & 0.2 & 0.25 & 0.33 & 0.39 & 0.7 & 0.7 & 0.58 & 0.8 & 0.8 & 0.85 & 0.36 & 0.43 \\
 \midrule
N\_M2\_L1 & 0.4 & 0.43 & \textbf{0.69} & 0.76 & 0.58 & 0.58 & 0.88 & 0.88 & \textbf{0.71} & 0.87 & \textbf{0.94} & \textbf{0.95} & 0.64 & 0.67 \\
N\_M2\_L2 & 0.47 & 0.55 & 0.58 & 0.76 & 0.58 & 0.57 & 0.88 & 0.89 & 0.6 & 0.75 & 0.93 & 0.94 & 0.67 & 0.67 \\
N\_M2\_L3 & 0.42 & \textbf{0.62} & 0.58 & 0.61 & 0.63 & 0.63 & 0.85 & 0.87 & 0.52 & 0.65 & 0.91 & 0.92 & 0.66 & 0.66 \\
N\_M2\_L4 & 0.25 & 0.26 & 0.52 & 0.54 & 0.46 & 0.49 & 0.81 & 0.84 & 0.52 & 0.65 & 0.76 & 0.76 & 0.5 & 0.53 \\
N\_M2\_L5 & 0.29 & 0.32 & 0.45 & 0.45 & 0.45 & 0.48 & 0.78 & 0.83 & 0.43 & 0.62 & 0.62 & 0.62 & 0.48 & 0.51 \\
N\_M2\_L6 & 0.14 & 0.13 & 0.22 & 0.26 & 0.3 & 0.35 & 0.67 & 0.7 & 0.29 & 0.57 & 0.52 & 0.52 & 0.3 & 0.36 \\
N\_M2\_L7 & 0.17 & 0.16 & 0.24 & 0.27 & 0.34 & 0.39 & 0.6 & 0.7 & 0.35 & 0.61 & 0.6 & 0.6 & 0.33 & 0.41 \\
 \midrule
N\_M3\_L1 & 0.35 & 0.35 & 0.62 & 0.7 & 0.5 & 0.53 & 0.9 & 0.9 & 0.63 & 0.88 & 0.91 & 0.92 & 0.58 & 0.62 \\
N\_M3\_L2 & 0.42 & 0.44 & 0.59 & 0.72 & 0.55 & 0.57 & 0.91 & 0.92 & 0.68 & 0.86 & \textbf{0.94} & \textbf{0.95} & 0.69 & 0.69 \\
N\_M3\_L3 & 0.31 & 0.3 & 0.23 & 0.27 & 0.44 & 0.51 & 0.9 & 0.91 & 0.63 & 0.85 & \textbf{0.94} & \textbf{0.95} & 0.65 & 0.67 \\
N\_M3\_L4 & 0.24 & 0.23 & 0.21 & 0.26 & 0.43 & 0.5 & 0.91 & 0.92 & 0.61 & 0.85 & 0.93 & 0.94 & 0.46 & 0.56 \\
N\_M3\_L5 & 0.23 & 0.23 & 0.21 & 0.25 & 0.43 & 0.5 & \textbf{0.93} & \textbf{0.93} & 0.65 & 0.85 & 0.93 & 0.94 & 0.47 & 0.57 \\
N\_M3\_L6 & 0.17 & 0.19 & 0.2 & 0.25 & 0.23 & 0.32 & 0.76 & 0.77 & 0.48 & 0.69 & 0.47 & 0.48 & 0.36 & 0.45 \\
N\_M3\_L7 & 0.18 & 0.19 & 0.21 & 0.25 & 0.19 & 0.31 & 0.75 & 0.76 & 0.53 & 0.77 & 0.63 & 0.63 & 0.31 & 0.37 \\
 \midrule
N\_M4\_L1 & 0.43 & 0.46 & 0.65 & 0.69 & 0.56 & 0.59 & 0.91 & 0.91 & 0.64 & \textbf{0.89} & \textbf{0.94} & 0.94 & 0.63 & 0.66 \\

N\_M4\_L2 & 0.42 & 0.44 & 0.62 & \textbf{0.78} & \textbf{0.74} & \textbf{0.73} & 0.91 & 0.92 & 0.65 & 0.84 & 0.84 & 0.85 & 0.74 & 0.74 \\

N\_M4\_L3 & \textbf{0.48} & 0.55 & 0.63 & 0.65 & 0.71 & 0.71 & 0.92 & \textbf{0.93} & 0.65 & 0.85 & 0.91 & 0.92 & \textbf{0.8} & \textbf{0.81} \\

N\_M4\_L4 & 0.33 & 0.33 & 0.5 & 0.51 & 0.57 & 0.59 & 0.83 & 0.86 & 0.62 & 0.82 & 0.78 & 0.79 & 0.64 & 0.65 \\
N\_M4\_L5 & 0.45 & 0.61 & 0.65 & 0.69 & \textbf{0.74} & \textbf{0.73} & 0.87 & 0.88 & 0.6 & 0.84 & 0.86 & 0.87 & 0.72 & 0.71 \\
N\_M4\_L6 & 0.13 & 0.13 & 0.55 & 0.72 & 0.19 & 0.3 & 0.67 & 0.7 & 0.46 & 0.68 & 0.59 & 0.59 & 0.24 & 0.32 \\
N\_M4\_L7 & 0.08 & 0.12 & 0.56 & 0.61 & 0.2 & 0.3 & 0.67 & 0.68 & 0.47 & 0.68 & 0.67 & 0.67 & 0.25 & 0.33 
\\
\bottomrule

\end{tabular}

\caption{
Summary of the performance of best-performing model-label combinations
of each \ac{NLI}-based model across each dataset}
\label{tab:RQ2N}
\end{table}

\begin{figure} [hbtp]
 \center
 \includegraphics[width=.98
 \textwidth]{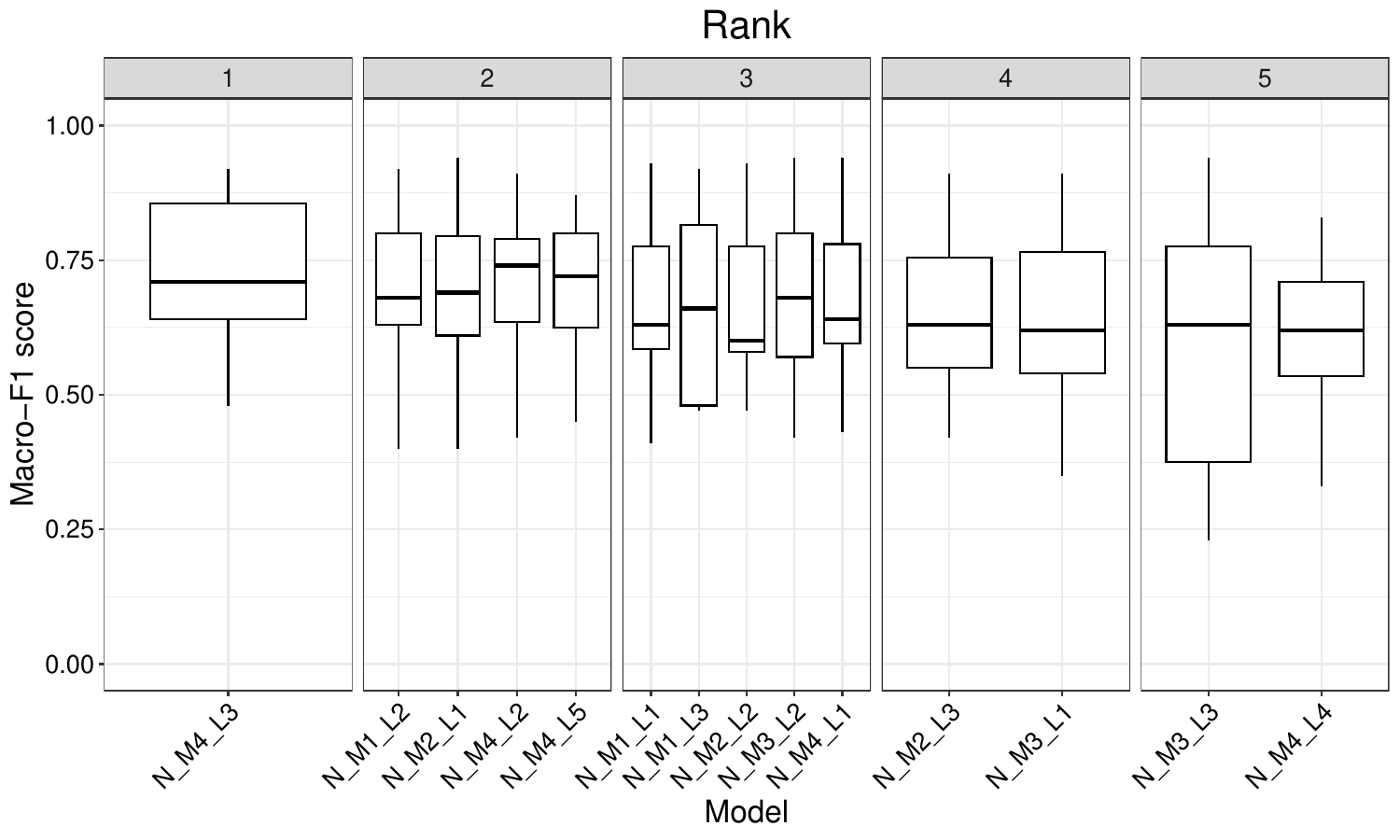}
 \caption{Scott-Knott ESD ranking for \ac{NLI}-based model-label combinations based on macro-F1 score}
 \label{fig:RQ2_NL}
\end{figure}

 Figure \ref{fig:RQ2_NL} presents the results of the Scott-Knott ESD test, where only the top 5 ranks are included due to space limitations. The results demonstrate the superiority of the N\_M4\_L3 combination, which was ranked first among all \ac{NLI}-based combinations. Both the combinations of N\_M4 with L2 and L5 labels were ranked second, along with the N\_M1\_L2 and N\_M2\_L1 combinations. The remaining combinations were distributed across the remaining ranks.

%%%%%%%%%%%%%%%%%%%%%%%%%%%%%%%
The results of combining \ac{TARS} model with different label configurations are presented in Table \ref{tab:RQ2T}, with the highest macro-F1 and micro-F1 score for each dataset are bolded. The results revealed the superiority of L2, which was able to produce the highest macro-F1 score for four datasets. L1 followed closely, achieving the highest scores for three datasets. The other labels failed to produce the highest values for any dataset.

\begin{table}[htbp]
\centering
\scriptsize
\begin{tabular}{@{}lcccccccccccccc@{}}
\toprule
\textbf{M\&L} & \multicolumn{2}{c}{\textbf{API}} & \multicolumn{2}{c}{\textbf{Gerrit}} & \multicolumn{2}{c}{\textbf{GitHub}} & \multicolumn{2}{c}{\textbf{Gitter}} & \multicolumn{2}{c}{\textbf{Google}} & \multicolumn{2}{c}{\textbf{Jira}} & \multicolumn{2}{c}{\textbf{\ac{SO}}} \\ 
 & \multicolumn{2}{c}{\textbf{reviews}} & \multicolumn{2}{c}{} & \multicolumn{2}{c}{} & \multicolumn{2}{c}{} & \multicolumn{2}{c}{\textbf{Play}} & \multicolumn{2}{c}{} & \multicolumn{2}{c}{\textbf{}} \\ 
 \midrule
 & \textbf{Mac} & \textbf{Mic} & \textbf{Mac} & \textbf{Mic} & \textbf{Mac} & \textbf{Mic} & \textbf{Mac} & \textbf{Mic} & \textbf{Mac} & \textbf{Mic} & \textbf{Mac} & \textbf{Mic} & \textbf{Mac} & \textbf{Mic} \\ 
 \midrule

T\_M1\_L1 & \textbf{0.49} & 0.63 & \textbf{0.64} & \textbf{0.7} & 0.49 & 0.52 & 0.77 & 0.78 & 0.59 & 0.67 & \textbf{0.89} & \textbf{0.9} & 0.62 & 0.62 \\
T\_M1\_L2 & 0.36 & \textbf{0.68} & 0.55 & 0.58 & \textbf{0.5} & \textbf{0.54} & \textbf{0.83} & \textbf{0.84} & \textbf{0.6} & \textbf{0.68} & 0.82 & 0.84 & \textbf{0.63} & \textbf{0.63} \\

T\_M1\_L3 & 0.33 & 0.46 & 0.52 & 0.59 & 0.44 & 0.44 & 0.62 & 0.62 & 0.58 & 0.65 & 0.6 & 0.64 & 0.36 & 0.36 \\
T\_M1\_L4 & 0.3 & 0.39 & 0.5 & 0.64 & 0.38 & 0.41 & 0.43 & 0.46 & 0.47 & 0.53 & 0.5 & 0.57 & 0.3 & 0.3 \\
T\_M1\_L5 & 0.3 & 0.38 & 0.48 & 0.6 & 0.34 & 0.38 & 0.42 & 0.43 & 0.22 & 0.22 & 0.4 & 0.5 & 0.3 & 0.31 \\
T\_M1\_L6 & 0.3 & 0.53 & 0.41 & 0.41 & 0.26 & 0.3 & 0.34 & 0.36 & 0.21 & 0.23 & 0.38 & 0.55 & 0.25 & 0.37 \\
T\_M1\_L7 & 0.3 & 0.64 & 0.54 & \textbf{0.7} & 0.3 & 0.33 & 0.37 & 0.37 & 0.32 & 0.39 & 0.4 & 0.63 & 0.27 & 0.33
\\

\bottomrule 

\caption{Summary of the performance of the \ac{TARS} model-label combinations
 across each dataset}
\label{tab:RQ2T}
\end{tabular}
\end{table}

\begin{figure} [hbtp]
 \center
 \includegraphics[width=.98
 \textwidth]{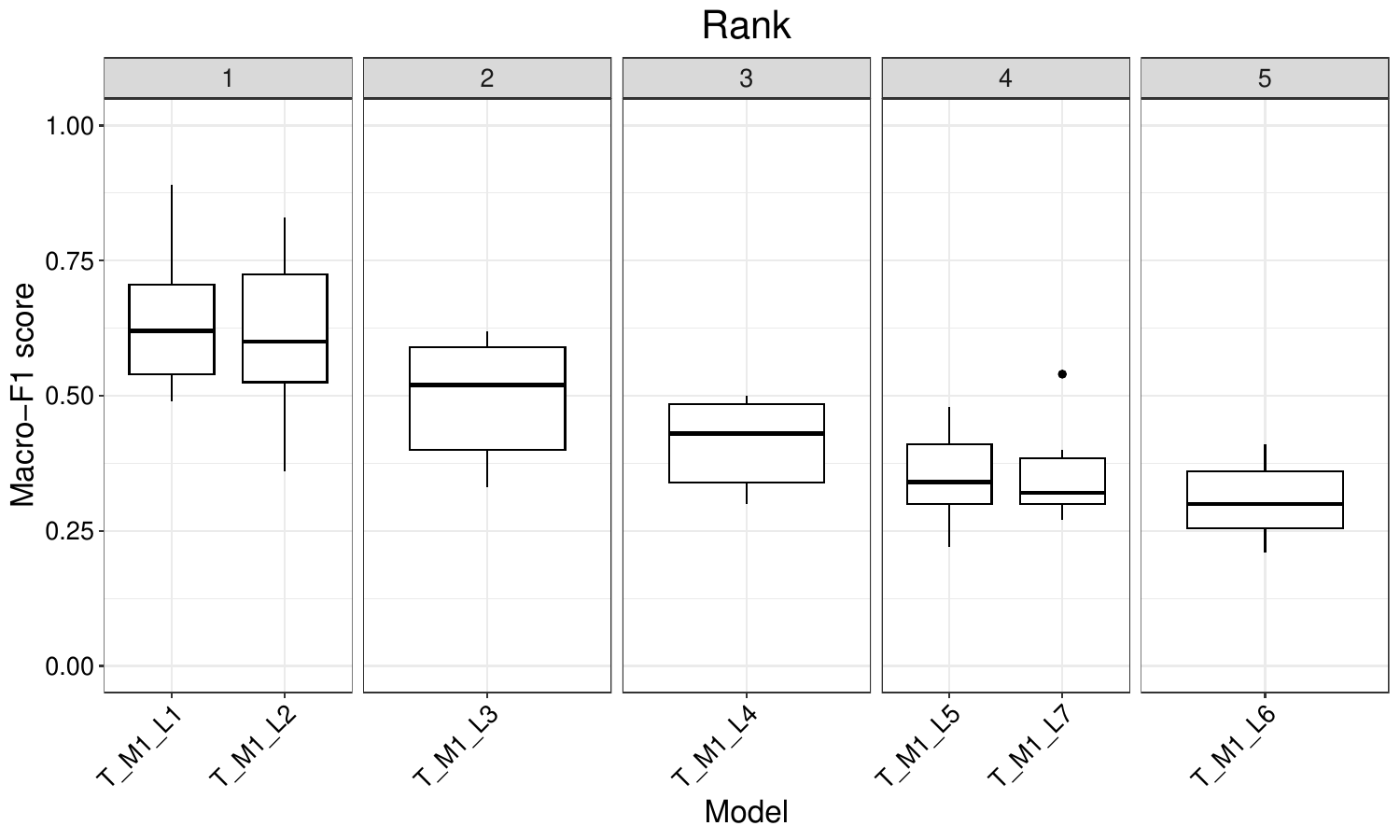}
 \caption{Scott-Knott ESD ranking for the \ac{TARS} model-label combinations based on macro-F1 score}
 \label{fig:RQ2_T}
\end{figure}

The results of the statistical test, presented in Figure \ref{fig:RQ2_T}, confirm these findings. The model combinations with L1 and L2 were ranked as the top performers, followed by L3 in second, L4 in third, L5 and L7 in fourth, and L6 in fifth.

Table \ref{tab:RQ2E} presents the results of combining the generative-based model with the label configurations, with the highest macro-F1 and micro-F1 score for each dataset are bolded. The results demonstrate the superiority of L1, which achieved the highest performance on three datasets. All other labels, except for L3, were able to achieve the highest value on one dataset, while L3 did not produce the highest value on any dataset.

\begin{table}[htbp]
\centering
\scriptsize
\begin{tabular}{@{}lcccccccccccccc@{}}
\toprule
\textbf{M\&L} & \multicolumn{2}{c}{\textbf{API}} & \multicolumn{2}{c}{\textbf{Gerrit}} & \multicolumn{2}{c}{\textbf{GitHub}} & \multicolumn{2}{c}{\textbf{Gitter}} & \multicolumn{2}{c}{\textbf{Google}} & \multicolumn{2}{c}{\textbf{Jira}} & \multicolumn{2}{c}{\textbf{\ac{SO}}} \\ 
 & \multicolumn{2}{c}{\textbf{reviews}} & \multicolumn{2}{c}{} & \multicolumn{2}{c}{} & \multicolumn{2}{c}{} & \multicolumn{2}{c}{\textbf{Play}} & \multicolumn{2}{c}{} & \multicolumn{2}{c}{\textbf{}} \\ 
 \midrule
 & \textbf{Mac} & \textbf{Mic} & \textbf{Mac} & \textbf{Mic} & \textbf{Mac} & \textbf{Mic} & \textbf{Mac} & \textbf{Mic} & \textbf{Mac} & \textbf{Mic} & \textbf{Mac} & \textbf{Mic} & \textbf{Mac} & \textbf{Mic} \\ 
 \midrule

G\_M1\_L1 & \textbf{0.52} & 0.58 & \textbf{0.75} & 0.8 & 0.73 & 0.73 & 0.9 & 0.91 & \textbf{0.67} & \textbf{0.89} & 0.88 & 0.89 & 0.72 & 0.73 \\

G\_M1\_L2 & 0.5 & 0.57 & 0.72 & 0.77 & 0.73 & 0.73 & 0.88 & 0.89 & 0.66 & \textbf{0.89} & 0.76 & 0.77 & \textbf{0.76} & \textbf{0.76} \\

G\_M1\_L3 & 0.5 & 0.57 & 0.72 & 0.77 & 0.64 & 0.65 & 0.86 & 0.88 & 0.62 & 0.86 & 0.82 & 0.82 & 0.72 & 0.73 \\
G\_M1\_L4 & 0.51 & 0.63 & \textbf{0.75} & \textbf{0.82} & 0.82 & 0.82 & 0.91 & 0.92 & 0.62 & 0.87 & 0.8 & 0.8 & 0.74 & 0.74 \\
G\_M1\_L5 & 0.38 & \textbf{0.64} & 0.62 & 0.8 & \textbf{0.83} & \textbf{0.83} & 0.9 & 0.9 & 0.62 & 0.86 & 0.82 & 0.83 & 0.62 & 0.63 \\
G\_M1\_L6 & 0.27 & 0.24 & 0.61 & 0.76 & 0.4 & 0.48 & 0.91 & 0.91 & 0.6 & 0.74 & \textbf{0.89} & \textbf{0.9} & 0.75 & 0.75 \\

G\_M1\_L7 & 0.22 & 0.22 & 0.69 & 0.8 & 0.32 & 0.38 & \textbf{0.93} & \textbf{0.93} & 0.6 & 0.86 & 0.81 & 0.82 & 0.64 & 0.66 \\ \bottomrule

\end{tabular}
\caption{Summary of the performance of the generative model-label combinations
across each dataset}
\label{tab:RQ2E}
\end{table}

\begin{figure} [hbtp]
 \center
 \includegraphics[width=.98
 \textwidth]{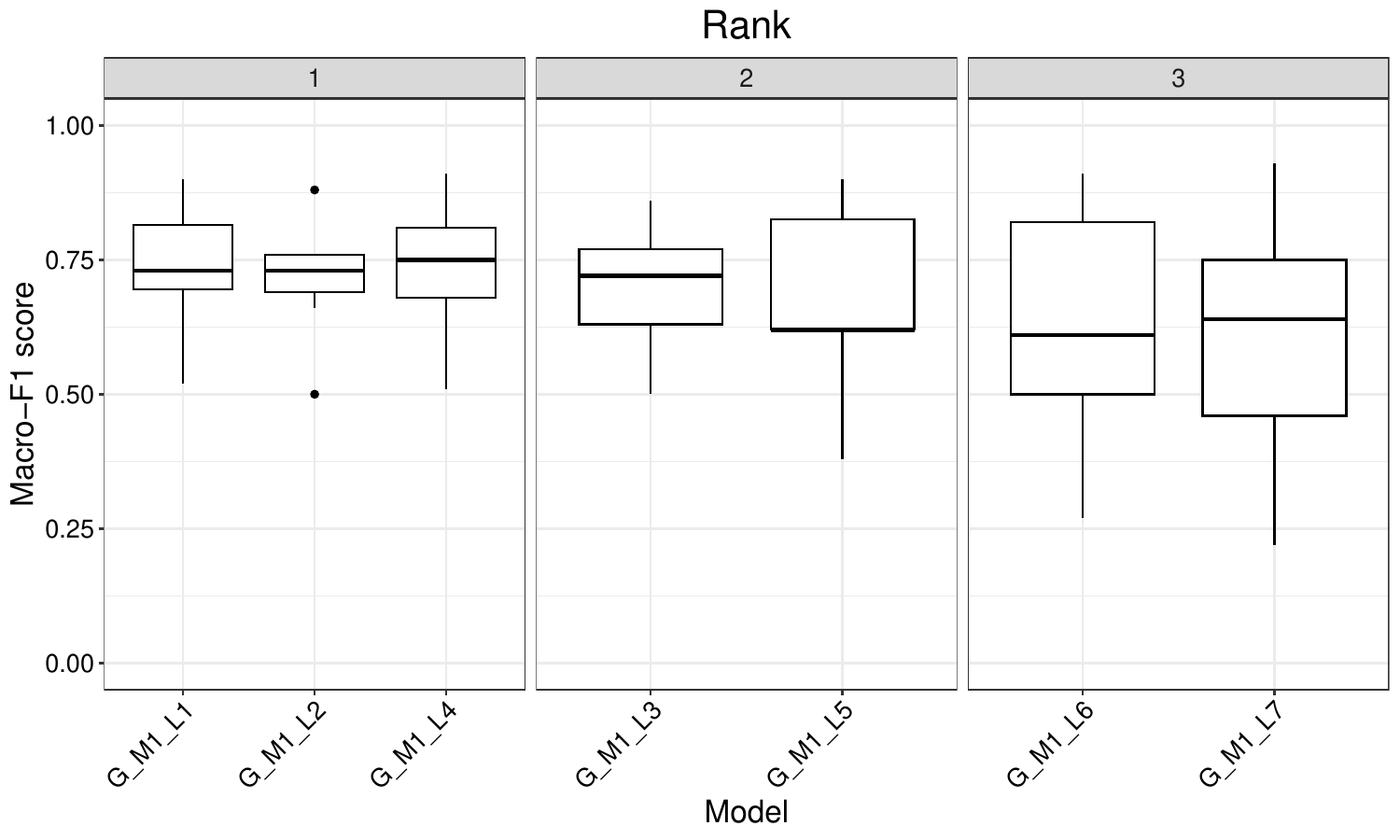}
 \caption{Scott-Knott ESD ranking for the generative model-label combinations based on macro-F1 score}
 \label{fig:RQ2_G}
\end{figure}

The statistical test produced three ranks, as Figure \ref{fig:RQ2_G} presents. Combining G\_M1 with L1, L2, or L4 resulted in the top performance. Pairing the model with L3 and L5 placed it in the second rank, while L6 and L7 combinations placed the model in the third rank.

%%%% Graph 

\begin{figure} [hbtp]
 \center
 \includegraphics[width=.98
 \textwidth]{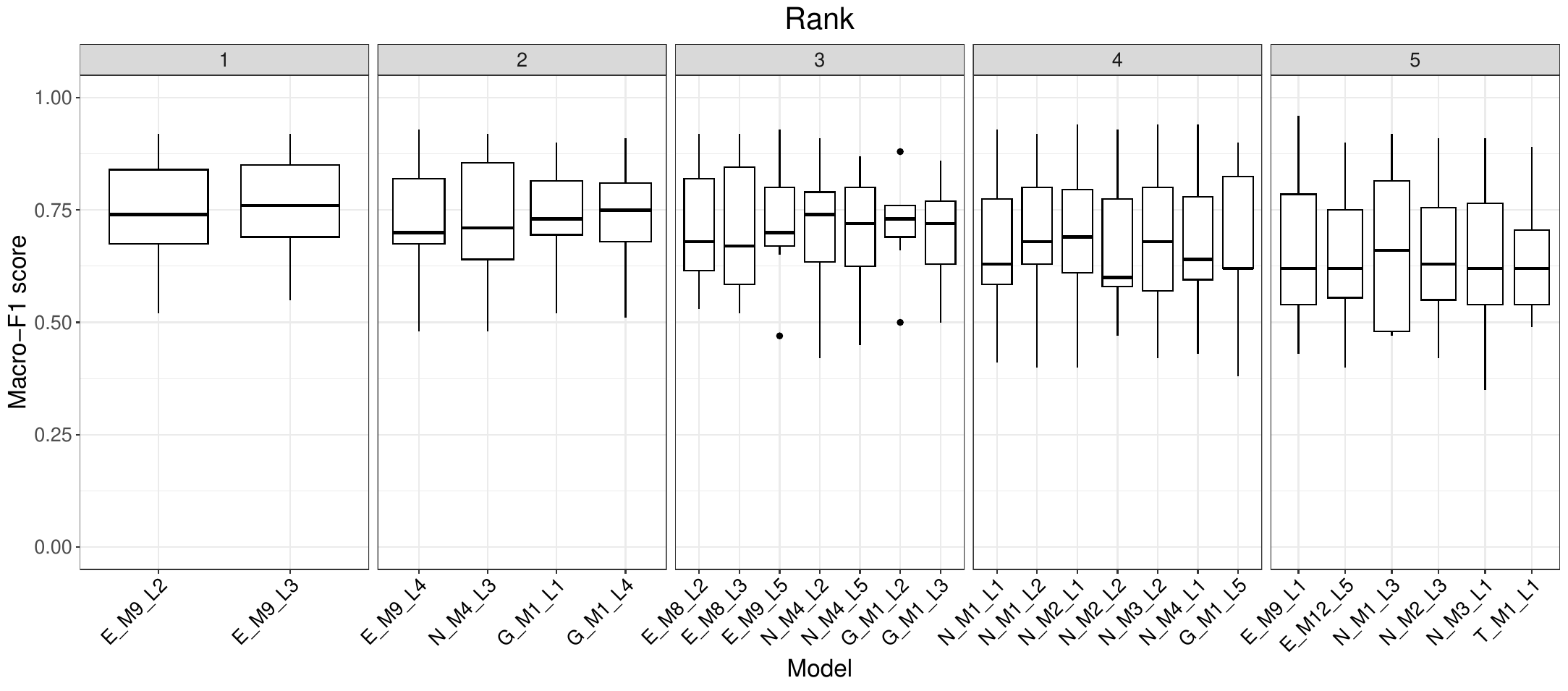}
 \caption{Scott-Knott ESD ranking for model-label combinations based on macro-F1 score}
 \label{fig:RQ2_All}
\end{figure}

The results of comparing the performance of all model-label combinations across all \ac{ZSL} techniques using the Scott-Knott ESD test are presented in Figure \ref{fig:RQ2_All}, with only the top 5 ranks included due to space limitations. The results confirm the superior performance of the E\_M9 model. Specifically, the model was ranked as the top performer when combined with both L2 and L3. Combining the model with L4 placed it in second place, alongside the generative-based model when combined with both L1 and L4. Additionally, the \ac{NLI}-based model (i.e., N\_M4) was ranked second when combined with L3. Other combinations of embedding, \ac{NLI}, and generative-based models followed in subsequent ranks. Notably, only one \ac{TARS}-based combination was ranked fifth, with other \ac{ZSL} technique combinations surpassing it.

\subsection{RQ3: How does the performance of \ac{ZSL}-based models compare with that of the state-of-the-art fine-tuned transformer-based models in sentiment classification?}

The results of the state-of-the-art fine-tuned transformer-based models and the best-performing model-label combinations for each \ac{ZSL}-based technique are presented in Table \ref{tab:RQ3}, where the highest values of the macro-F1 and micro-F1 scores for each dataset are bolded.

As the table depicts, F\_M2\_L1 achieved the highest macro scores in 3 out of the 7 datasets, and G\_M1\_L4 attained the highest scores in 2 datasets. Furthermore, F\_M1\_L1, F\_M3\_L1, and E\_M9\_L1 each achieved the highest score in one dataset.

\begin{figure} [hbtp]
 \center
 \includegraphics[width=.98
 \textwidth]{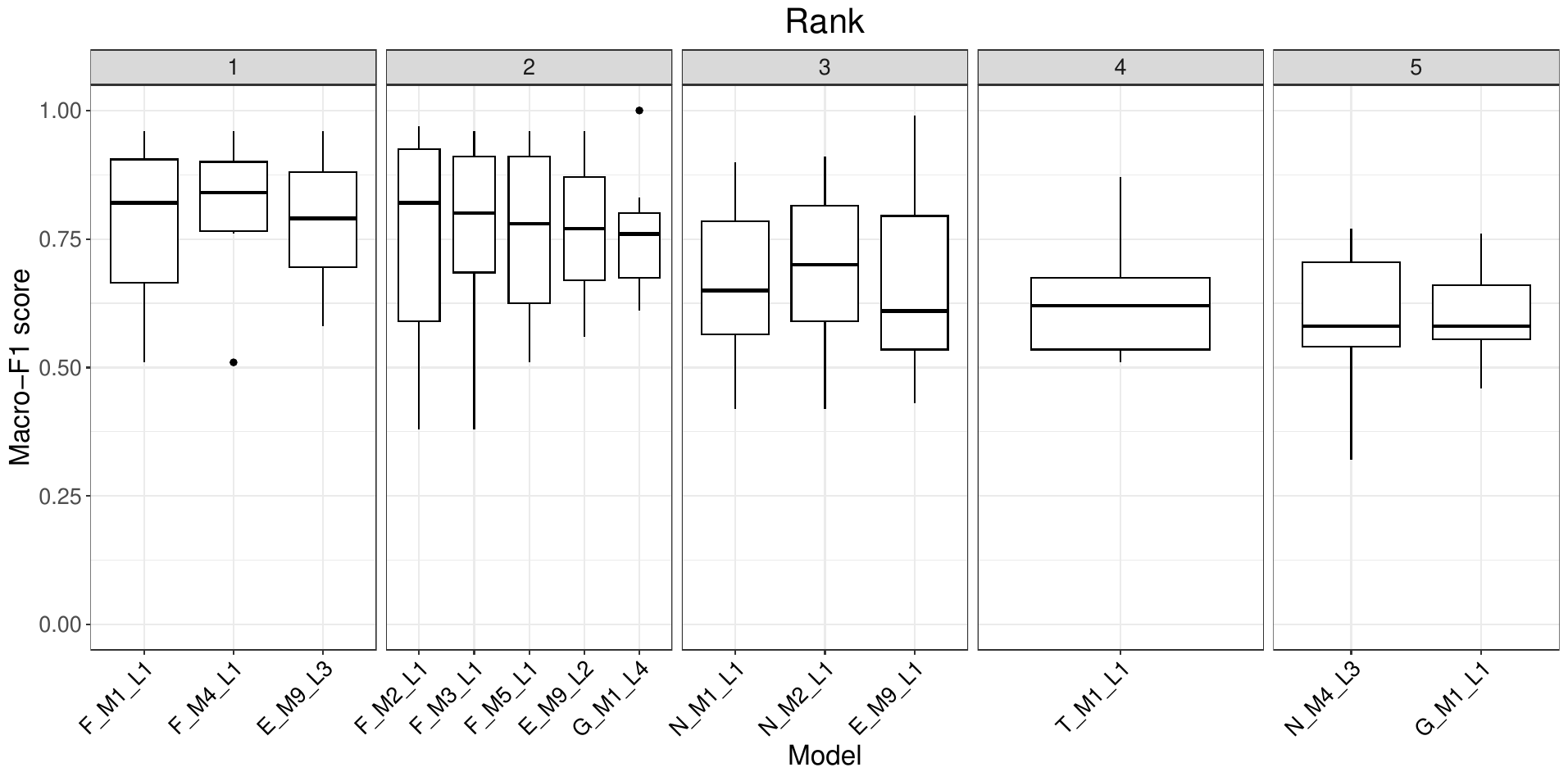}
 \caption{Scott-Knott ESD ranking for the state-of-the-art fine-tuned transformer-based models and the best-performing model-label combinations for each \ac{ZSL}-based technique based on macro-F1 score}
 \label{fig:RQ3}
\end{figure}

The results of the statistical analysis are presented in Figure \ref{fig:RQ3}. The statistical analysis revealed that F\_M1\_L1 , F\_M4\_L1 , and E\_M9\_L3 are the top performers. The remaining fine-tuned models along with E\_M9\_L2 and G\_M1\_L4 , were ranked second. N\_M1\_L1, N\_M2\_L1, and E\_M9\_L1 were ranked third. T\_M1\_L1 preceded N\_M4\_L3 and G\_M1\_L1 , with the latter two models ranked last, in the fifth rank.

The above results revealed that embedding-based and generative-based \ac{ZSL} models have competitive performance compared to that of fine-tuned models on the testing set. 

\begin{table}[htbp]
\centering
\scriptsize
\begin{tabular}{@{}lcccccccccccccc@{}}
\toprule
\textbf{M\&L} & \multicolumn{2}{c}{\textbf{API}} & \multicolumn{2}{c}{\textbf{Gerrit}} & \multicolumn{2}{c}{\textbf{GitHub}} & \multicolumn{2}{c}{\textbf{Gitter}} & \multicolumn{2}{c}{\textbf{Google}} & \multicolumn{2}{c}{\textbf{Jira}} & \multicolumn{2}{c}{\textbf{\ac{SO}}} \\ 
 & \multicolumn{2}{c}{\textbf{reviews}} & \multicolumn{2}{c}{} & \multicolumn{2}{c}{} & \multicolumn{2}{c}{} & \multicolumn{2}{c}{\textbf{Play}} & \multicolumn{2}{c}{} & \multicolumn{2}{c}{\textbf{}} \\ 
 \midrule
 & \textbf{Mac} & \textbf{Mic} & \textbf{Mac} & \textbf{Mic} & \textbf{Mac} & \textbf{Mic} & \textbf{Mac} & \textbf{Mic} & \textbf{Mac} & \textbf{Mic} & \textbf{Mac} & \textbf{Mic} & \textbf{Mac} & \textbf{Mic} \\ 
 \midrule

 \multicolumn{15}{c}{{Embedding-based \ac{ZSL} models} } \\ 
 \midrule
E\_M9\_L1 & 0.47 & 0.49 & 0.43 & 0.75 & 0.61 & 0.62 & 0.95 & 0.95 & 0.6 & 0.86 & \textbf{0.99} & \textbf{0.99} & 0.64 & 0.67 \\
E\_M9\_L2 & 0.56 & 0.67 & 0.79 & 0.84 & 0.74 & 0.73 & 0.95 & 0.95 & 0.6 & 0.86 & 0.96 & 0.97 & 0.77 & 0.77 \\
E\_M9\_L3 & 0.58 & 0.67 & 0.79 & 0.86 & 0.79 & 0.78 & 0.95 & 0.95 & 0.6 & 0.86 & 0.96 & 0.97 & 0.81 & 0.81 \\
 \midrule
\multicolumn{15}{c}{\ac{NLI}-based \ac{ZSL} models} \\ \midrule 
N\_M1\_L1 & 0.42 & 0.44 & 0.67 & 0.73 & 0.55 & 0.57 & 0.9 & 0.9 & 0.58 & 0.83 & 0.9 & 0.91 & 0.65 & 0.68 \\
N\_M2\_L1 & 0.42 & 0.45 & 0.77 & 0.83 & 0.58 & 0.59 & 0.86 & 0.86 & 0.6 & 0.86 & 0.91 & 0.92 & 0.7 & 0.72 \\
N\_M4\_L3 & 
0.32 & 0.33 &
0.53 & 0.54 & 
0.58 & 0.6 & 
0.77 & 0.81 &
0.56 & 0.8 & 
.76 & 0.76 &
0.65 & 0.66\\
\midrule 
\multicolumn{15}{c}{\ac{TARS}-based \ac{ZSL} models} \\ \midrule 
T\_M1\_L1 & 0.55 & 0.65 & 0.62 & 0.69 & 0.51 & 0.53 & 0.71 & 0.71 & 0.52 & 0.69 & 0.87 & 0.89 & 0.64 & 0.63 \\
 \midrule 
\multicolumn{15}{c}{{Generative-based \ac{ZSL} models} } \\ \midrule 
G\_M1\_L1 & 0.57 & 0.69 & 0.66 & 0.69 & 0.46 & 0.54 & 0.66 & 0.68 & 0.58 & 0.83 & 0.54 & 0.81 & 0.76 & 0.76 \\
G\_M1\_L4 & 0.51 & 0.68 & 0.76 & 0.83 & 0.77 & 0.77 & \textbf{1} & \textbf{1} & \textbf{0.61} & \textbf{0.89} & 0.83 & 0.84 & 0.74 & 0.73 \\
 \midrule
 
\multicolumn{15}{c}{Fine-tuned models} \\ \midrule

F\_M1\_L1 & \textbf{0.82} & \textbf{0.89} & 0.78 & 0.85 & 0.92 & 0.92 & 0.51 & 0.67 & 0.55 & 0.8 & 0.96 & 0.97 & 0.89 & 0.89 \\
F\_M2\_L1 & 0.75 & 0.84 & \textbf{0.82} & \textbf{0.87} & \textbf{0.94} & \textbf{0.94} & 0.38 & 0.62 & 0.43 & 0.63 & 0.97 & 0.98 & \textbf{0.91 }& \textbf{0.91} \\
F\_M3\_L1 & 0.8 & 0.87 & 0.79 & 0.84 & \textbf{0.94} & \textbf{0.94} & 0.38 & 0.62 & 0.58 & 0.83 & 0.96 & 0.97 & 0.88 & 0.88 \\
F\_M4\_L1 & 0.77 & 0.86 & 0.76 & 0.82 & 0.92 & 0.92 & 0.84 & 0.86 & 0.51 & 0.74 & 0.96 & 0.97 & 0.88 & 0.88 \\
F\_M5\_L1 & 0.72 & 0.81 & 0.78 & 0.84 & 0.93 & 0.93 & 0.51 & 0.67 & 0.53 & 0.77 & 0.96 & 0.97 & 0.89 & 0.89 \\

\bottomrule
\end{tabular}
\caption{
Summary of the performance of state-of-the-art fine-tuned transformer-based models and the best-performing model-label combinations for each \ac{ZSL}-based technique}
\label{tab:RQ3}
\end{table}

\subsection{RQ4: What factors contribute to the misclassification of sentiment labels in \ac{ZSL}-based models, and how do these compare with those shared with the state-of-the-art fine-tuned transformer-based models?}

The results of the quantitative error analysis are summarized in Table \ref{tab:RQ4}, where the lowest number of misclassified instances for each dataset are bolded. Among the \ac{ZSL}-based models,  E\_M9\_L3 yielded the fewest errors across the majority of datasets (i.e., 3 out of 7 datasets), followed by G\_M1\_L1 and G\_M1\_L4, in 2 out of 7 datasets, and E\_M9\_L1 in only one dataset. 
Moreover, common misclassifications across the \ac{ZSL}-based models included 35 API reviews, 19 GitHub comments, 3 app reviews, 1 Jira comment, and 10 \ac{SO} posts.

When comparing the performance of the fine-tuned models to the \ac{ZSL}-based models, the fine-tuned models produced fewer misclassifications than all of the examined \ac{ZSL}-based models in 4 out of 7 datasets. The fine-tuned models were able to reduce misclassifications on API reviews, Gerrit code review comments, GitHub comments, and \ac{SO} posts. The common misclassifications by the fine-tuned models included 25 API reviews, 11 Gerrit code review comments, 16 GitHub comments, 1 Gitter developer message, 5 app reviews, 1 Jira comment, and 25 \ac{SO} posts.

Both \ac{ZSL}-based models and fine-tuned models failed to classify 9 API reviews, 5 GitHub comments, 3 app reviews, 1 Jira comment, and 1 \ac{SO} post.

The analysis of the commonly misclassified instances by the \ac{ZSL}-based models revealed that among the total 68 misclassified instances, 64.71\% were originally annotated as neutral, 22.06\% as positive, and 13.24\% as negative. 

The categorization of errors indicated that subjectivity in annotation resulting from different annotators’ perceptions of emotions accounts for 60.29\% of the commonly misclassified instances. Polar facts were identified as the second most common source of misclassification, where it represents 22.06\% of the errors. Polar facts are those that describe inherently desirable or undesirable situations, expressed in a neutral tone, such as the comment ``this doesn't work".
Politeness errors accounted for 8.82\% of the misclassified instances. These errors arose when polite expressions, such as ``Thanks!", led to inconsistent classifications. Misclassifications due to figurative language, such as humor, irony, or sarcasm, contributed to 4.41\% of the errors. An example is the API review statement: ``So initializing high is better than too low", which was annotated as negative by human annotators but incorrectly classified as positive or neutral by the models. Finally, pragmatic errors, such as statements reporting third-party opinions or emotions, were another source of misclassification, accounting for 4.41\% of the errors. In these cases, sentences that humans identify as neutral are misclassified as positive or negative by the models due to the presence of emotion-related words. An example of this is the API review statement: ``I know many developers would like to have this", which was annotated as neutral, but the majority of models classified it as positive.

The common misclassified instances among both \ac{ZSL}-based and fine-tuned models were originally labeled as follows: 57.89\% neutral, 26.32\% positive, and 15.79\% negative. The error categorization revealed that subjectivity in annotation was the main contributor, accounting for 73.68\% of these instances. Politeness errors accounted for 15.79\%, followed by polar facts with 5.26\%, and figurative language with 5.26\%.

\begin{table}[htbp]
\scriptsize
\caption{Summary of missclassifieds instances}
\label{tab:RQ4}
\begin{tabular}{@{}llllllll@{}}
\toprule
\textbf{M\&L} & \textbf{API } & \textbf{Gerrit} & \textbf{GitHub} & \textbf{Gitter} & \textbf{Google} & \textbf{Jira} & \textbf{\ac{SO}} \\

 & \textbf{reviews} & & & & \textbf{Play} & & 
\\

 \midrule
 
Test set size & 452 & 160 & 713 & 21 & 35 & 93 & 443 \\
 \midrule
 \multicolumn{8}{c}{\ac{ZSL}-based models} \\
 \midrule

E\_M9\_L1 & 229 & 40 & 270 & 1 & 5 & \textbf{1} & 148 \\
E\_M9\_L2 & 148 & 25 & 193 & 1 & 5 & 3 & 101 \\
E\_M9\_L3 & 149 & 23 & 154 & 1 & 5 & 3 & 86 \\

N\_M1\_L1 & 254 & 43 & 308 & 2 & 6 & 8 & 143 \\
N\_M2\_L1 & 250 & 27 & 294 & 3 & 5 & 7 & 126 \\

N\_M4\_L3 & 304 & 74
& 285 & 4 & 7 

& 22 & 151 \\

T\_M1\_L1 & 157 & 50 & 333 & 6 & 11 & 10 & 164 \\

G\_M1\_L1 & 138 & 50 & 330 & \textbf{0} & 6 & 18 & 105 \\
G\_M1\_L4 & 143 & 27 & 162 & \textbf{0} & \textbf{4} & 15 & 118 \\
 \midrule

Common among \ac{ZSL}-based models & 35 & 0
& 19 & 0 & 3 &

1 & 10\\

 \midrule
 \multicolumn{8}{c}{Fine-tuned models} \\
 \midrule
F\_M1\_L1 & \textbf{51} & 24 & 54 & 7 & 7 & 3 & 47 \\
F\_M2\_L1 & 74 & \textbf{21} & \textbf{42} & 8 & 13 & 2 & \textbf{41} \\
F\_M3\_L1 & 60 & 25 & 50 & 8 & 6 & 3 & 51 \\
F\_M4\_L1 & 62 & 27 & 57 & 3 & 9 & 3 & 52 \\
F\_M5\_L1 & 85 & 26 & 52 & 7 & 8 & 3 & 50 \\

 \midrule
Common among fine-tuned models & 25 & 11 & 16 & 1 & 5 & 1 & 25 \\
 \midrule
 \midrule

Common among all & 9 & 0 & 5 & 0 & 3 & 1 & 1 \\
\bottomrule
\end{tabular}
\end{table}

\section{Discussion}\label{Discussion} 
%% comapre with previous work
The results of RQ1 revealed that generative-based \ac{ZSL} outperformed other \ac{ZSL} techniques using original labels, with \ac{NLI} ranked second. Despite being pre-trained for sentiment analysis in tweets, E\_M9 performed well on software engineering sentiment analysis. Although, the paid embedding (i.e., E\_M10) outperformed most embeddings, it was outperformed by some \ac{NLI}-based models and E\_M9. Therefore, we recommend exploring freely available models before investing in a paid one, as the former may yield better results.

RQ2 revealed that no single label configuration consistently outperformed others across all models. While the original label L1 performed well with TARS and generative-based \ac{ZSL}, embedding-based and \ac{NLI}-based models performed better with expert-curated labels, particularly those incorporating the sentiment term (i.e., L2 and L3). Combining E\_M9 with L3 yielded the best results, suggesting that including task context in label configurations may enhance performance. Additionally, combining E\_M9 with L3 yielded the best results across all \ac{ZSL} techniques. This supports the earlier observation that pre-trained models from other domains are a viable approach for \ac{ZSL}.

In RQ3, when compared to fine-tuned state-of-the-art models, E\_M9 paired with L3 ranked as the top performer, achieving results comparable to some fine-tuned models, while surpassing others. This finding contradicts previous research, which observed that general sentiment analysis tools underperform in software engineering contexts \cite{jongeling2015choosing,tourani2014monitoring}, as the finding demonstrates that pre-trained models trained on sentiment analysis in other domains can achieve performance similar to or exceeding fine-tuned models without the need for additional training or fine-tuning, thereby addressing the issue of data scarcity.

RQ4 revealed that most misclassifications in \ac{ZSL}-based models occurred in the neutral class, consistent with the observations of \cite{lin2022opinion} and \cite{obaidi2021development} that neutral sentiments are challenging to classify. Subjectivity in annotation and polar facts were identified as the primary causes of these misclassifications. Subjectivity in annotation was also highlighted by \cite{obaidi2021development} as one of the challenges in sentiment analysis.

Another observation is that while RQ3 demonstrated that some \ac{ZSL}-based models performed similarly to fine-tuned models, the latter models were more effective at reducing misclassifications in technical datasets (i.e., API reviews, code reviews, GitHub comments, and \ac{SO} posts), as evidenced by the results of RQ4. In contrast, \ac{ZSL}-based models performed better on more conversational datasets, such as developer chat messages, app reviews, and Jira comments.

When comparing our results with previous, related studies, our results can be compared with the work of \cite{zhang2020sentiment}, which was performed in RQ3. The other work that can be compared to ours is the work of \cite{zhang2023revisiting}.
Specifically, we compare the performance of our \ac{ZSL}-based models with their \ac{ZSL}-based and \ac{FSL}-based models on the common datasets from both studies. A key difference between our study and \cite{zhang2023revisiting} is that while we evaluated the models on all test sets in RQ3 and RQ4, \cite{zhang2023revisiting} selected a stratified representative random sample due to the high cost of running generative models.

In our study, combining E\_M9 with both L2 and L3 resulted in a 0.79 macro-F1 score on the Gerrit dataset, higher than the 0.76 best macro-F1 score achieved by \cite{zhang2023revisiting} through \ac{FSL}, where their \ac{ZSL} achieved 0.75. 
For GitHub, our highest macro-F1 score is 0.79, achieved by E\_M9\_L3, while \cite{zhang2023revisiting} achieved 0.72 with both \ac{FSL} and \ac{ZSL}. \cite{zhang2023revisiting} achieved a perfect macro-F1 score on the Google Play dataset with \ac{FSL} and 0.98 with \ac{ZSL}, whereas our best \ac{ZSL} model (i.e., G\_M1\_L4) achieved only 0.61. E\_M9\_L1 achieved an almost perfect macro-F1 score (i.e., 0.99) on the Jira dataset, where \cite{zhang2023revisiting} achieved 0.91 through \ac{FSL} and 0.85 through \ac{ZSL}.

Although \cite{zhang2023revisiting} focused solely on generative-based models, the comparison highlights the superiority of E\_M9, which achieved higher results on 3 out of the 4 common datasets. This confirms our observation that applying \ac{ZSL} with a model pre-trained on a similar context can yield competitive performance.

%%%%%RQ4

\section{Threats to validity} \label{Threats to validity} 
Several threats may affect the validity of this study. This section discusses external, construct, internal, and conclusion validity threats, along with corresponding mitigation strategies, where applicable \cite{shull2007guide}.

\subsection{External validity}
 A potential threat to our study's external validity is the generalizability of its findings. Although our results are derived from seven sentiment classification datasets representing diverse software engineering contexts, the generalizability of these findings cannot be claimed beyond these datasets. 

Another potential external validity threat arises from the models and labels utilized. Since our findings are confined to the models and labels examined, they may not generalize to others that were not included in this study. Consequently, we acknowledge this as a limitation to our study's external validity. 

\subsection{Construct validity}
A potential threat to this study's construct validity lies in the performance measures employed. To address this threat, we report both macro-F1 and micro-F1 scores, with the former being the basis for comparison. This approach aligns with the approach of \cite{novielli2020can,zhang2023revisiting}.

Another potential construct validity threat stems from the annotation of the utilized datasets. We rely on datasets annotated in prior work and widely utilized in related studies \cite{lin2018sentiment,zhang2020sentiment,zhang2023revisiting}. While efforts have been made to ensure annotations accuracy, we cannot guarantee the complete absence of errors. Consequently, this limitation is inherited from the datasets.

An additional potential construct validity threat is the manual categorization of misclassified instances, which may introduce subjectivity. To mitigate this threat, we independently categorized the errors then resolved inconsistencies through discussion until a consensus was reached.

\subsection{Internal validity}
A potential threat to our study's internal validity is the possibility of implementation errors. To mitigate this, we used widely accepted and validated model implementations and followed guidelines and tutorials for applying these models in similar contexts, such as \cite{alammar2024hands,tunstall2022natural}.

\subsection{Conclusion validity}
Conclusion validity concerns the appropriate use of statistical tests \cite{shull2007guide}. Besides using common statistical measures, such as percentages, to convey the study’s findings, we only employed Cohen’s Kappa coefficient to measure inter-rater agreement and the non-parametric Scott-Knott ESD test to compare the performance of models. 

The use of Cohen’s Kappa coefficient aligns with its intended purpose and is recommended by empirical standards in software engineering research \cite{ralph2021empirical}. The Scott-Knott ESD test was selected for its ability to produce disjoint groups with non-negligible magnitudes of difference, its application in prior studies within similar contexts, and its high tolerance for outliers \cite{tantithamthavorn2018impact, puth2015effective}.

\section{Conclusion} \label{Conclusion} % Reem

Aiming to address the challenge of data scarcity within sentiment analysis for software engineering, this study explored the potential of \ac{ZSL} for sentiment analysis in software engineering. We empirically evaluated the performance of embedding-based, \ac{NLI}-based, \ac{TARS}-based, and generative-based \ac{ZSL} techniques across multiple software engineering contexts. We expanded the scope of our analysis by including various label configurations to understand their impact on model performance. We then compared the best-performing \ac{ZSL}-based models and label combinations with the state-of-the-art fine-tuned transformer-based models. Finally, we conducted an error analysis to better understand the causes of sentiment misclassifications by \ac{ZSL}-based models.

The results demonstrated that \ac{ZSL} is a viable approach for sentiment analysis within the software engineering domain, achieving performance comparable to or exceeding state-of-the-art fine-tuned transformer-based models without requiring fine-tuning. Specifically, an embedding-based model, fine-tuned to analyze sentiments within tweets, achieved top performance when paired with expert-generated labels that contextualized the dataset and incorporated the term ``sentiment”. The combination of the model and label was statistically ranked as the top performer, where it either matched or surpassed fine-tuned models. Similarly, the generative-based model when paired with expert-generated labels that elaborated on describing emotions with sentiments was ranked as the second-best performer alongside some of the fine-tuned models. The error analysis revealed that most misclassified instances were originally annotated as neutral sentiment, and the further categorization of the misclassified instances identified subjectivity in annotation and polar facts as the primary causes of misclassification.

These findings underscore the potential of \ac{ZSL} in addressing the challenges of training data shortages for sentiment analysis in software engineering. By eliminating the need for annotated data, \ac{ZSL} provides a practical solution for sentiment analysis across diverse software engineering contexts.

Future work could explore how linguistic variation in label phrasing impacts sentiment classification performance. Additionally, investigating the influence of different prompting strategies on model effectiveness in sentiment classification represents a promising avenue for future research. Moreover, a deeper error analysis that leverages explainable artificial intelligence (XAI) and examines differences between developer-oriented and user-generated datasets could provide more nuanced insights into \ac{ZSL} performance within sentiment analysis in software engineering.

\section*{Declaration of Generative AI and AI-assisted Technologies in the Writing Process}
The authors used ChatGPT to improve the readability and language of the manuscript, with full responsibility for the final content.

\bibliographystyle{elsarticle-num}

\bibliography{bibliography}

\end{document}